\documentclass[prd,aps,showpacs,nofootinbib,preprintnumbers,twocolumn]{revtex4-1}
\usepackage{graphicx,graphics,epsfig,epstopdf,subfigure,color,psfrag}% Include figure files
\usepackage{hyperref}
\newcommand{\be}{\begin{equation}}
\newcommand{\ee}{\end{equation}}
\newcommand{\bea}{\begin{eqnarray}}
\newcommand{\eea}{\end{eqnarray}}
\newcommand{\nn}{\nonumber}
\newcommand{\dd}{\displaystyle}

\begin{document}

\preprint {BARI-TH/2014-687}

\title{Rare semileptonic $B\to K^* \ell^+ \ell^- $ decays  in RS$_c$ model}
\author{P.~Biancofiore$^{a,b}$, P.~Colangelo$^{a}$,  F.~De~Fazio$^{a}$}
\affiliation{
$^{a}$INFN, Sezione di Bari, via Orabona 4, I-70126 Bari, Italy\\
$^{a}$Dipartimento di Fisica, Universit\'a  di Bari,via Orabona 4, I-70126 Bari, Italy\\}
\begin{abstract}
Recent LHCb measurements show small discrepancies  with respect to the Standard Model (SM) predictions in  selected angular distributions of the mode $B^0 \to K^{*0} \mu^+ \mu^-$. The possibility of explaining such  tensions  within theories beyond the SM crucially depends on the size of the deviations of the Wilson coefficients of the effective Hamiltonian for this mode,  in comparison to their SM values. We analyse this issue in the framework of the Randall Sundrum model with custodial protection (RS$_c$);   in our study  we also consider the  mode with $\tau$ leptons in the final state.
We discuss the small deviations of RS$_c$ results from SM ones, found scanning the  parameter space of the model. 
\end{abstract}
\pacs{1260.Cn,1260.Fr,1320.He}
\maketitle

\section{Introduction}
Despite the  phenomenological success of the Standard Model (SM),  there are hints that the paths to new physics are  open, and that they pass across the flavour sector.  Joint efforts at the $B$ factories and at the hadron colliders have provided us with data of unprecedented precision in this sector,  so that we are now sensitive to small effects in  theoretical calculations that are essential in the comparison with the measurements. Moreover, the large sample of collected data  has  further enriched the experimental scenario, and information is now available on particular processes  forseen as the most promising ones to unveil deviations from SM.
Among those, a prominent role is played by the measurement of the  rare  $B_{s,d} \to \mu^+ \mu^-$ branching fractions \cite{Aaij:2013aka,Chatrchyan:2013bka}:
\bea
{\bar {\cal B}}(B_s \to \mu^+ \mu^-)&=& (2.9 \pm 0.7) \times 10^{-9} \,\,\,\, ,\nn\\
{\cal B} (B_d \to \mu^+ \mu^-)&=&(3.6 \pm^{1.6}_{1.4}) \times 10^{-10} \,\,\,\,\,\, , \label{bsdmumuexp}
\eea
where the symbol $\bar {\cal B}$, in the case of $B_s$,  indicates that the effect of width difference in the $B_s - \bar B_s$ system is taken into account \cite{Dunietz:2000cr}. The corresponding  SM predictions  are \cite{Bobeth:2013uxa}:
\bea
{\bar {\cal B}}(B_s \to \mu^+ \mu^-)_{SM}&=&(3.65 \pm 0.23) \times 10^{-9}\,\,\,\, ,\nn\\
{\cal B}(B_d \to \mu^+ \mu^-)_{SM}&=&(1.06 \pm 0.09) \times 10^{-10}\label{bsdmumuSM} \,\,,
\eea
 including the errors on the  meson lifetimes, on the top quark mass and $\alpha_s$, on the values of the CKM matrix elements and on the decay constants $f_{B_{s,d}}$.  The comparison between (\ref{bsdmumuexp}) and (\ref{bsdmumuSM}) shows the existing  agreement in  the case of  $B_s$, which excludes  large contributions from new scalar or pseudoscalar particles;  in the case of  $B_d \to \mu^+ \mu^-$ the SM prediction undershoots the measurement,  but the uncertainties are too large to draw conclusions.

Another process recognized as particularly sensitive to new physics effects is the rare semileptonic $B \to K^* \mu^+ \mu^-$ decay, mainly due to the numerous observables that can be studied to disentangle  additional  new particle contributions in this  loop-driven transition. Several measurements were already available from $B$ factories, such as the branching fraction, the  forward-backward lepton asymmetry and the $K^*$ longitudinal polarization fraction in few bins of   the squared  momentum  transferred to the lepton pair, $q^2$.  Analyses at LHC have improved the scenario, enlarging the set of the measured observables \cite{Aaij:2011aa,Aaij:2013iag,Aaij:2013qta,Usanova:2013osa,Chatrchyan:2013cda}. A recent LHCb investigation has reported  a set of 24 measurements, disclosing a problematic effect in  particular observables, as described  below. This has prompted several discussions aimed at explaining the discrepancy and at understanding in which direction new physics effects should  prominently appear.

In this work, we  study the  $B \to K^* \ell^+ \ell^-$ mode in the framework of the Randall-Sundrum model,
a new physics framework justified by several theoretical considerations \cite{RandallSundrum}. In this model,   the  space-time  is supposed to be five-dimensional with  warped metric, so that in its 4D reduction new particles appear that are either Kaluza-Klein excitation of SM particles, or  new ones having no SM counterparts.   Additional contributions  to the considered rare mode are therefore produced, and we  discuss whether they can explain the observed pattern of measurements, preserving the agreement of the computed $\bar {\cal B}(B_s \to \mu^+ \mu^-)$ with data.
In our analysis we consider both the cases $\ell=\mu$ (and $e$) and $\ell=\tau$. The reason is that tensions with SM predictions have also been   found in the branching fractions of semileptonic 
$B$ decays with a $\tau$ lepton in the final state \cite{Lees:2012xj,taus}. It is therefore interesting to work out  predictions for other decay modes involving the $\tau$ lepton.

Several studies of $B$ decays in the RS framework exist in the literature \cite{burdman,agashe,moreau,neubertRS,buras1,buras2,Albrecht:2009xr,buras4,Blanke:2012tv}. 
An important issue concerns the mass scale of the lowest Kaluza-Klein excitations $M_{KK}$ that is compatible with existing data on flavour observables.
In this respect the main problem (the so-called flavour problem in RS)  is represented by the parameter $\epsilon_K$ describing indirect CP violation in the Kaon sector. Agreement with data for this parameter would require   large $M_{KK}$, of the  order of tens TeV. In order to allow mass values  in a range accessible to  the LHC without   a strong fine tuning  to predict $\epsilon_K$ in its experimental   range, several solutions  have been proposed,  namely  an extended gauge group for strong interactions \cite{ Bauer:2011ah} or additional symmetries \cite{Santiago:2008vq,Fitzpatrick:2007sa}. In the framework of the so-called custodial RS$_c$ model, 
it has been shown that it is possible to have $M_{KK}= {\cal O}(1 \,\,{\rm TeV})$ 
%of the order of a few TeV 
without requiring too much fine tuning to reproduce $\epsilon_K$ in the measured range  \cite{buras1,buras2,Albrecht:2009xr,buras4}. For such low mass scales,  sizeable deviations from SM predictions in $\Delta F=2$ observables are possible, with the main role in  particle-antiparticle mixing diagrams coming from the first KK excitation of the gluon, at least in the Kaon sector, while for $B_{d,s}$ mixing also KK excitations of the electroweak gauge bosons play a role,  and  in particular the new gauge boson $Z_H$,  that we introduce below, is important  \cite{burdman,buras1}. On the other hand,  it has been shown that  the effects are less pronounced in the inclusive $\Delta F=1$ $B$ decays, at odds with the Kaon $\Delta F=1$  transitions  \cite{buras2}. In both sectors, the main new contribution comes from the modification of $Z$ couplings to right-handed down-type quarks, since the left-handed ones are protected by the custodial symmetry. 

In \cite{buras2} the inclusive $b \to s \ell^+ \ell^-$ decay modes were considered neglecting the modifications of the coefficients of the dipole operators in the relevant effective Hamiltonian  which appear at loop level. The changes of such coefficients in the RS framework have been computed to study the inclusive $B \to X_s \gamma$ decay mode  \cite{Blanke:2012tv}.
Our analysis is complementary to those performed in \cite{buras2,Blanke:2012tv} in several respects,  since for example we study the exclusive mode  $B \to K^* \ell^+ \ell^-$, which requires   a set of hadronic form factors.
Furthermore, we  provide a calculation of the coefficients of the dipole operators using a different technique with respect to that carried out in \cite{Blanke:2012tv}, which more consistently matches with  the other parts of the calculation.  A prime attention is paid to the  bounds provided to the  parameter space  of the model by the present  experimental knowledge, in particular of the CKM mixing matrix. Moreover, we study in details the modes with $\tau$ leptons, and the sensitivity of their  additional observables to this new physics scenario.

The plan of the paper is the following. In Sec.~\ref{sec:bkmumu} we fix our notations for the effective Hamiltonian governing the process $B \to K^* \ell^+ \ell^-$, and collect the definitions  required in the subsequent discussion.
In Sec.~\ref{sec:RS} we describe the main features of the RS$_c$ model, in particular the particle content and the couplings relevant for our study.  The numerical analysis follows in Sec.~\ref{sec:num} and \ref{sec:results}, where we investigate all the measured observables in $B \to K^* \mu^+ \mu^-$  and present a set of predictions for the mode with $\tau$ leptons. Concluding remarks can be found in Sec.~\ref{sec:conc}. In the Appendices  we collect several technical details, in particular those relevant for our calculation of the Wilson coefficients $\Delta C_{7,8}^{(\prime)}$: the profiles of  various KK states, the couplings of neutral bosons to fermions,  and the different contributions to the Wilson coefficients.

\section{General features of the $B \to K^{*} \ell^+ \ell^-$ decay mode }\label{sec:bkmumu} 
The  effective $ \Delta B =-1$, $\Delta S = 1$ Hamiltonian governing  the rare transition $b \to s \ell^+ \ell^-$ can be written as
\bea
H^{eff}&=&-\,4\,{G_F \over \sqrt{2}} V_{tb} V_{ts}^* \, \Big\{C_1 O_1+C_2 O_2 \nn \\
&+&\sum_{i=3,..,6}C_i O_i+\sum_{i=7,..,10,P,S} \left[ C_i O_i +C_i^{\prime  } O_i^{\prime  } \right] \Big\}\, , \,\,\,\,\,\, \label{hamil}
\eea
where $G_F$ is the Fermi constant and $V_{ij}$ are
elements of the Cabibbo-Kobayashi-Maskawa mixing matrix. We
neglect doubly Cabibbo suppressed terms proportional to $V_{ub} V_{us}^*$. $O_1$ and $O_2$ are current-current operators, and $O_i$ $(i=3,\dots, 6)$ are QCD penguins;  they have a minor impact on the modes  we are considering,  therefore we neglect them
and refer to \cite{Buchalla:1995vs} for details about their structure.
Among the remaining operators, the primed ones have opposite chirality with respect to the unprimed. Only the unprimed ones, for $i=7,\dots10$, are present in the SM;  the scalar $O_S$ and pseudoscalar $O_P$ operators, although present in SM, are highly  suppressed.
 %and  can be neglected  as well. 
 Therefore,
 the operators  considered in our analysis are 
\begin{itemize}
\item Magnetic penguin operators:
\bea
O_7&=&{e \over 16 \pi^2} m_b ({\bar s}_{L \alpha} \sigma^{\mu \nu}
     b_{R \alpha}) F_{\mu \nu} \nn \\
O_7^\prime &=& \frac{e}{16 \pi^2}m_b ({\bar s}_{R \alpha} \sigma^{\mu \nu}  b_{L \alpha}) F_{\mu \nu}  \nn \\
O_8&=&{g_s \over 16 \pi^2} m_b \Big[{\bar s}_{L \alpha} \sigma^{\mu \nu} \Big({\lambda^a \over 2}\Big)_{\alpha \beta} b_{R \beta}\Big] \;
      G^a_{\mu \nu}  \label{mag-peng} \\
      O_8^\prime&=&{g_s \over 16 \pi^2} m_b \Big[{\bar s}_{R \alpha} \sigma^{\mu \nu} \Big({\lambda^a \over 2}\Big)_{\alpha \beta} b_{L \beta}\Big] \;
      G^a_{\mu \nu} \,\, ,\nonumber
      \eea
 \item Semileptonic electroweak penguin operators:
\bea
O_9&=&{e^2 \over 16 \pi^2}  ({\bar s}_{L \alpha} \gamma^\mu b_{L \alpha}) \; {\bar \ell} \gamma_\mu \ell  \nn \\
O_9^\prime& =&{e^2 \over 16 \pi^2}  ({\bar s}_{R \alpha} \gamma^\mu b_{R \alpha}) \; {\bar \ell} \gamma_\mu \ell  \nn \\
O_{10}&=&{e^2 \over 16 \pi^2}  ({\bar s}_{L \alpha} \gamma^\mu b_{L \alpha}) \; {\bar \ell} \gamma_\mu \gamma_5 \ell  \label{eq-peng} \\  
O_{10}^\prime&=&{e^2 \over 16 \pi^2}  ({\bar s}_{R \alpha} \gamma^\mu b_{R \alpha}) \; {\bar \ell} \gamma_\mu \gamma_5 \ell\,\,.\nonumber
\eea
\end{itemize}
In (\ref{mag-peng}-\ref{eq-peng})  $\alpha$ and $\beta$ are colour indices, $\lambda^a$ the Gell-Mann matrices,  and  
$\displaystyle b_{R,L}={1 \pm \gamma_5 \over 2}b$.  
$F_{\mu \nu}$ and $G^a_{\mu \nu}$  denote the
electromagnetic and the gluonic field strength tensors, respectively, and $e$ and $g_s$ are the
electromagnetic and the strong coupling constants. $m_b$ is the $b$ quark mass, while   the operators proportional to the strange quark mass $m_s$ are neglected.

Considering the subsequent resonant $K^* \to K \pi$ decay, the $B \to K^*( \to K \pi) \ell^+ \ell^-$ fully differential decay width   can be written as follows:
\be
\frac{d^4 \Gamma (B \to K^*[\to K \pi] \ell^+ \ell^-) }{dq^2 d\cos \theta_\ell d\cos \theta_K d\phi}=
\frac{9}{32 \pi}I(q^2,\theta_\ell, \theta_K, \phi) \,\,\, , \label{fully-diff}
\ee
where
\bea
I(q^2,\theta_\ell, \theta_K, \phi) &=& I_1^s \sin^2 \theta_K+I_1^c \cos^2 \theta_K \nn \\
&+&(I_2^s \sin^2 \theta_K+I_2^c \cos^2 \theta_K)\cos 2 \theta_\ell \nn \\
&+&I_3 \sin^2 \theta_K \sin^2\theta_\ell \cos 2 \phi \nn \\
&+&I_4 \sin 2 \theta_K \sin 2 \theta_\ell \cos \phi \nn \\
&+&I_5 \sin 2 \theta_K \sin  \theta_\ell \cos \phi \label{fully-diff-1}\\\
&+&(I_6^s \sin^2 \theta_K +I_6^c \cos^2 \theta_K)\cos \theta_\ell \nn \\
&+& I_7 \sin 2 \theta_K \sin \theta_\ell \sin \phi \nn \\
&+&I_8 \sin 2 \theta_K \sin 2 \theta_\ell \sin \phi \nn \\
&+&I_9 \sin^2 \theta_K \sin^2 \theta_\ell \sin 2 \phi \,. \nn
\eea
In the previous equations, $q^2$ is the dilepton  invariant mass; $\theta_K$ is  the angle between the Kaon
direction  and the direction opposite to  the $B$ meson one  in  the $K^*$ rest frame; $\theta_\ell$ is
 the angle between the charged lepton direction\footnote{LHCb uses the $\ell^+$ direction for $B^0$ decays,  $\ell^-$  for $\bar B^0$.} and  the direction opposite to that
 the $B$ meson  in the  lepton pair rest frame;  finally, $\phi$ is  the angle between the plane containing the lepton pair and the plane containing the $K^*$ decay products, i.e. $K$ and $\pi$
 \cite{Faessler:2002ut,Altmannshofer:2008dz}. 
 In  the case of the CP conjugated mode, the $\bar B$ meson decay, one defines analogous functions $\bar I$ in which all the weak phases are conjugated.   The fully  differential decay width 
$d \bar \Gamma$ can be written in analogy to (\ref{fully-diff}), with the function $I$ replaced by $\bar I$ which is obtained from (\ref{fully-diff-1}) by the  rule \cite{Altmannshofer:2008dz}:
\bea
I_{1,2,3,4,7} &\to& {\bar I}_{1,2,3,4,7}  \nn \\
I_{5,6,8,9} &\to& -{\bar I}_{5,6,8,9} \,\,\,.
\eea
All the functions $I_i(\bar I_i)$ can be written in terms of eight transversity amplitudes, which in turn are functions of the $B \to K^*$ form factors.
We adopt the following parametrization for the $B \to K^*$ matrix elements:
\bea
<K^*(p^\prime,\epsilon)|{\bar s} \gamma_\mu (1-\gamma_5) b| B(p)>=\hspace*{3.5cm}  \nn \\
 \epsilon_{\mu \nu \alpha \beta} \epsilon^{* \nu} p^\alpha p^{\prime \beta}
{ 2 V(q^2) \over M_B + M_{K^*}}  \hspace*{0.5cm} \nn \\
- i \left [ \epsilon^*_\mu (M_B + M_{K^*}) A_1(q^2) -
(\epsilon^* \cdot q) (p+p')_\mu  {A_2(q^2) \over M_B + M_{K^*} }
\right. \nn\\
- \left. (\epsilon^* \cdot q) {2 M_{K^*} \over q^2}
\big(A_3(q^2) - A_0(q^2)\big) q_\mu \right ] \,\, , \,\,\,  \hspace*{0.5cm} \label{a1}
\eea
\bea
<K^*(p^\prime,\epsilon)|{\bar s} \sigma_{\mu \nu} q^\nu
{(1+\gamma_5) \over 2} b |B(p)>= \hspace*{2cm} \nn \\
 i \epsilon_{\mu \nu \alpha
\beta} \epsilon^{* \nu} p^\alpha p^{\prime \beta}
\; 2 \; T_1(q^2)   \hspace*{0.5cm}  \nn \\
+  \Big[ \epsilon^*_\mu (M_B^2 - M^2_{K^*})  -
(\epsilon^* \cdot q) (p+p')_\mu \Big] \; T_2(q^2) \hspace*{0.5cm}  \nn \\
+ (\epsilon^* \cdot q) \left [ q_\mu - {q^2 \over M_B^2 -
M^2_{K^*}} (p + p')_\mu \right ] \; T_3(q^2)  \, , \,\,\, \,\,\, 
\label{t1}
\eea
where $\epsilon$ is the $K^*$ polarization vector.
The various form factors are not all independent;  $A_3$ can be written as
\be
A_3(q^2) = {M_B + M_{K^*} \over 2 M_{K^*}}  A_1(q^2) - {M_B -
M_{K^*} \over 2 M_{K^*}}  A_2(q^2)  
\ee
with  $A_3(0) = A_0(0)$; moreover, the identity 
$\displaystyle
\sigma_{\mu \nu} \gamma_5 = - {i \over 2}
 \epsilon_{\mu \nu \alpha \beta} \sigma^{\alpha \beta}$
(with $\epsilon_{0 1 2 3}=+1$) implies   $T_1(0) = T_2(0)$.
Using these definitions, the transversity amplitudes entering in the  $I_i$ structures,  and the $I_i$ functions  themselves, can be found in Ref.~\cite{Altmannshofer:2008dz} (their Eqs.~(3.28)-(3.45)), with the only change that the three $T_i$ form factors in that paper have to be divided by a factor of $2$ to match our definition in Eq.(\ref{t1}).

From the functions $I_i(q^2)$ and $\bar I_i(q^2)$, CP conserving quantities ($S_i$) and CP asymmetries ($A_i$) can be built:
\bea
S_i&=& \frac{I_i+\bar I_i}{ \frac{d\Gamma}{dq^2}+\frac{d\bar \Gamma}{dq^2} }\label{Si} \,\, , \\
A_i&=& \frac{I_i-\bar I_i}{ \frac{d\Gamma}{dq^2}+\frac{d\bar \Gamma}{dq^2}}\label{Ai} \,\,.
\eea
Starting from these quantities, several observables can be introduced. In particular, we  consider
\begin{itemize}
\item the lepton forward-backward (FB) asymmetry: $A_{FB}=-\frac{3}{8} (2S_6^s+S_6^c)$;
\item the longitudinal $K^*$ polarization fraction: $F_L=-S_2^c$;
\item binned observables $S_i$ , with their numerators and denominators separately integrated over $q^2$ bins of the kind $[q^2_1,q^2_2]$: $<S_i>_{[q^2_1,q^2_2]}$.
\end{itemize}
These are the main observables analyzed in the present  investigation, that will be compared to the experimental results. 

Of great interest is the lepton FB asymmetry. The first analyses  by BaBar, Belle and CDF  Collaborations  seemed to contradict the SM expectation for the sign of $A_{FB}(q^2)$ in the low $q^2$ region: $q^2 \in[1-6]$ GeV$^2$ \cite{Aubert:2008bi,Wei:2009zv,Aaltonen:2011ja}. The presently available improved  analysis   by LHCb shows  agreement with SM \cite{Aaij:2011aa}, as we discuss in more details below. 
On general grounds, $A_{FB}$ might or  not have  a zero in the kinematically accessible $q^2$ range. 
The position of such a zero, in terms of the $B \to K^*$  form factors and of the Wilson coefficients in the   effective Hamiltonian  (\ref{hamil}), is given by the value of $q^2$ for which  the  equation holds:
\bea
{\rm Re} \Bigg\{ \frac{2m_b}{q^2} \left[ (M_B+M_{K^*})\frac{T_1(q^2)}{V(q^2)}
(C_7+C_7^\prime)(C_{10}-C_{10}^\prime)^* \right. \nn \\
\left. +(M_B-M_{K^*})\frac{T_2(q^2)}{A_1(q^2)}(C_7-C_7^\prime)(C_{10}+ \,C_{10}^\prime)^*\right]\nn \\
+C_9\,C_{10}^* -C_9^\prime C_{10}^{\prime *}\Bigg\}=0 \,\,\,\,\, . \nn \\ \label{zero-gen}
\eea
In SM this  equation becomes:
\bea
{\rm Re} \left\{ \frac{2m_b}{q^2} \Big[ \Big((M_B+M_{K^*})\frac{T_1(q^2)}{V(q^2)} \right. \hspace*{3cm} \nn \\
\left. +(M_B-M_{K^*})\frac{T_2(q^2)}{A_1(q^2)}\Big)C_7 \,C_{10}^*\Big] +C_9\,C_{10}^* \right\}=0 \,\, .\,\,\, \,\,\,\,\,\, \label{zero-sm}
\eea
In the large energy limit of  $K^*$  and in the heavy quark limit  the $q^2$ dependence of the form factor ratios  in (\ref{zero-gen}), (\ref{zero-sm})  cancels out, and $T_1/V=M_B/(M_B+M_{K^*})$, $T_2/A_1=(M_B+M_{K^*})/M_B$ (modulo radiative corrections);
hence, the position of the zero of $A_{FB}$ is, to a large extent,  a  quantity which only depends on the structure of the interactions, i.e. it is independent  of the form factors,  and at NLO for the Wilson coefficients it is given by $q_0^2|_{SM}=4.39\pm^{0.38}_{0.35}$ GeV$^2$ \cite{Beneke:2001at}. This value must be compared with the LHCb determination: $q_0^2|_{LHCb}=4.9 \pm 0.9$ GeV$^2$ \cite{Aaij:2013iag}.

Results that have raised interest are those reported by the LHCb Collaboration \cite{Aaij:2013qta}, with the measurement of the observables  \cite{DescotesGenon:2012zf}
\be
P^\prime_{i=4,5,6,8}=\frac{S_{i=4,5,7,8}}{\sqrt{F_L(1-F_L)}}
\label{Piprime} \,\, 
\ee
related to $F_L$ and $S_i$ defined  above.
The measurement is carried out in $6$   bins of $q^2$ for each one of the four observables in (\ref{Piprime}):  a discrepancy is found in the case of $P_5^\prime$ in the third $q^2$ bin, where the datum is  significantly lower than the SM prediction, as we show below.
A small deviation is also found in $P_4^\prime$ for another value of $q^2$.
 Efforts have been devoted to identify the kind of new physics effects which may explain the full set of data without altering the observables in agreement with SM predictions. The general idea is to try to understand which one of the Wilson coefficients (and how many of them) should be modified (increased/suppressed),   including  those  not present in  SM,  to reproduce the data \cite{Descotes-Genon:2013wba,Altmannshofer:2013foa,Beaujean:2013soa}.

In the following we do not adopt the phenomenological approach of looking separately at the various Wilson coefficients, but rather we  make use of a specific new physics scenario,  the custodially protected Randall-Sundrum model, in which the weak effective Hamiltonian emerges from a well defined theory of 
elementary interactions. The resulting Wilson coefficients are therefore correlated, and such a correlation has precise phenomenological consequences to  be considered in the various observables, namely  those in (\ref{Piprime}).
 Moreover, 
motivated by the experimental results of semileptonic and leptonic $B$ decays to $\tau$ leptons, we also study   observables  for the case of massive final leptons, namely the $\tau$  polarization asymmetries.
 
To define polarization asymmetries,  let us consider the spin vector $s$ of  the $\tau^-$ lepton having momentum $k_1$, with  $s^2=-1$ and $k_1 \cdot s=0$.  In the  $\tau^-$ rest frame 
three orthogonal
unit vectors can be defined, $\vec e_L$, $\vec e_N$ and $\vec e_T$,  corresponding to the
longitudinal $s_L$, normal $s_N$ and transverse $s_T$ polarization vectors:
\bea 
s_L &=& (0,\vec e_L)=\left( 0, {\vec k_1 \over |\vec k_1|} \right) \nonumber \\
s_N &=& (0,\vec e_N)=\left( 0, {\vec p^\prime \times \vec k_1 \over |\vec p^\prime \times \vec k_1|} \right)  \label{spinsrf} \\
s_T &=& (0,\vec e_T)=(0,\vec e_N \times \vec e_L) \,\,\, .  \nonumber  
\eea
\noindent
In Eq.(\ref{spinsrf}) $\vec p^\prime$ and $\vec k_1$ are the
$K^* $  and  $\tau^-$  three-momenta in the rest frame of the lepton pair.  Choosing  the
$z$-axis directed as the $\tau^-$ momentum in the 
lepton pair rest frame we have $k_1=(E_1,0,0,|\vec k_1|)$, and boosting the spin vectors $s$
in (\ref{spinsrf}) in the rest frame of the lepton pair,  the normal and transverse polarization vectors
$s_N$, $s_T$ remain unchanged,  $s_N=(0,1,0,0)$ and $s_T=(0,0,-1,0)$, while the longitudinal polarization vector becomes:
\be 
s_L={1 \over m_\tau}(|\vec k_1|,0,0,E_1) \,\,\, . \label{sl} 
\ee
For each value of the squared momentum transfered to the lepton pair,  $q^2$,
the polarization asymmetry for the negatively charged  $\tau^-$ lepton  can be defined as: 
\be 
{\cal A}_A(q^2)={ \dd{  d \Gamma \over dq^2}(s_A)-{d\Gamma \over dq^2}(-s_A) \over { \dd {d \Gamma \over dq^2}(s_A)+{d
\Gamma\over dq^2} (-s_A)} }\,\,\, , \label{def-pol} 
\ee 
with $A=L$, $T$ and $N$.
These quantities have been analysed in the SM (e.g.  in \cite{Kruger:1996cv,Colangelo:2006gv} and in the references therein). 
In particular, in Ref.\cite{Colangelo:2006gv}  it has been pointed out that the $\tau$ polarization asymmetries are also form factor independent quantities in the $q^2$ region where the large energy limit can be applied.
These results can be  generalized including the new primed operators considered here. In particular, 
one can  exploit the expressions in \cite{Colangelo:2006gv} for the observables in $B \to K^* \tau^+ \tau^-$ with the  substitutions
\bea
C_7 \, T_1& \to & (C_7 + C_7^\prime) \, T_1 \nn \\
C_7 \, T_{2,3}& \to & (C_7 - C_7^\prime) \, T_{2,3} \nn \\
C_9 \, V& \to & ( C_9 +  C_9^\prime) \, V \nn \\
C_{10} \, V& \to & ( C_{10} +C_{10}^\prime) \, V \label{Kstar-RS} \\
C_9 \, A_{1,2}& \to & ( C_9 -  C_9^\prime) A_{1,2} \nn \\
C_{10} \, A_{1,2,0}& \to & ( C_{10} -  C_{10}^\prime) \, A_{1,2,0}
 \,\,. \nn
\eea
We adopt these prescriptions in the following discussion.

\section{Randall-Sundrum model with custodial protection}\label{sec:RS}

In this Section we describe the main features of the Randall-Sundrum (RS) model \cite{RandallSundrum}, a theoretical framework  constructed with the motivation, among others, represented by  the possibility of solving the  hierarchy problem  and of explaining the observed hierarchies in the fermion masses and mixing angles.
The model is defined in  a five dimensional space-time manifold with coordinates $(x,y)$  ($x$ the ordinary Minkowskian coordinates) and  metric
\bea
ds^2&=&e^{-2 k y} \eta_{\mu \nu} dx^\mu dx^\nu - dy^2 \,\,\, , \nn \\
\eta_{\mu \nu}&=&diag(+1,-1,-1,-1) \,\,\, . \label{metric}
\eea
The scale parameter $k$ is chosen $k\simeq {\cal O}(M_{Planck})$  to address the hierarchy problem; we set it to $k=10^{19}$ GeV. The (fifth) coordinate $y$ varies in a range between two branes, $0 \le y \le L$;  $y=0$ corresponds to the so-called UV brane,   $y=L$ to the IR one.

Several variants of the model have been proposed, each one adding new features to those of the original model. Here, we consider the  scenario in which 
the SM gauge symmetry  group is enlarged to the gauge group
\be
SU(3)_c \times SU(2)_L \times SU(2)_R \times U(1)_X \times P_{L,R} \,\,
\label{group}
\ee
which, together with the metric,   defines the Randall-Sundrum model with custodial protection RS$_c$ \cite{contino,carena,cacciapaglia}. 
The custodial protection is realized imposing the  discrete $Z_2$ $P_{L,R}$ symmetry,  which implies a mirror action of the two $SU(2)_{L,R}$ groups,
preventing large $Z$ couplings to left-handed fermions that would be incompatible with experiment. Furthermore, this variant has been proven to be consistent with electroweak precision observables for masses of the lightest Kaluza-Klein excitations of the order of a few TeV \cite{Contino:2006qr,Carena:2007ua},  in the reach of the LHC.

Two symmetry breakings occur:  first, the gauge group (\ref{group}) is broken to the SM gauge group imposing suitable boundary conditions (BC) on the UV brane. Afterwards, the spontaneous symmetry breaking occurs, which is Higgs-driven as in SM.
All the SM fields are allowed to propagate in the bulk, except for the Higgs field which is localized close to the IR brane. Here, we consider the case of a Higgs boson completely localized at $y=L$.

The extension of the SM  leads to the presence of new particles,  as a consequence of the requirements/assumptions   listed below.
\begin{itemize}
\item
The  two $SU(2)$ groups require a larger  number of gauge bosons. Those corresponding to $SU(2)_L$ are $W_L^{a,\mu}$  ($a=1,2,3$),  while $W_R^{a, \mu}$ correspond to $SU(2)_R$. The gauge conditions $W_{L,R}^{a,5}=0$ and $\partial_\mu W_{L,R}^{a,\mu}=0$ are chosen, as well as for all the other gauge bosons.
The $P_{L,R}$ symmetry imposes the equality $g_L=g_R=g$  for the  $SU(2)_{L,R}$ gauge couplings.
\par
The number of remaining gauge bosons is the same as in  SM.
In particular, the eight gauge fields corresponding to $SU(3)_c$ are still identified with the gluons, while     the $U(1)_X$ gauge field is denoted as $X_\mu$, with  coupling  $g_X$.
The 5D couplings are dimensionful: the relations to their 4D counterparts are
$g^{4D}={g^{5D}}/{\sqrt{L}}$.
We shall describe below  the mixing pattern among the various gauge fields.
\item
Fitting matter fields in suitable representations of the group (\ref{group}) leads to new fermions, as  discussed in the following. 
\item
The  presence of a compact fifth dimension implies the existence of a tower of Kaluza-Klein (KK) excitations for all particles. As generically done in extra-dimensional models, the boundary conditions  help to distinguish particles having a SM correspondent  from those without SM partners, by requiring the existence or not of a zero mode in the KK mode expansion of a given field.
Two choices for  BC are considered: Neumann BC  on both branes (++), or Dirichlet BC on the UV brane and Neumann BC on the IR one (-+). Only fields with  (++) BC have a zero mode which can be identified with a SM particle.
\end{itemize}

For each one of the fields   listed above we perform a KK decomposition of the generic form:
\be
F(x,y)=\frac{1}{\sqrt{L}}\sum_k F^{(k)}(x) f^{(k)}(y)\,\,,
\label{genericKK}
\ee
referring to the functions $f^{(k)}(y)$ (specific for each field $F$) as the 5D field profiles, while $F^{(k)}(x)$ are the corresponding effective 4D fields. Then, we consider the 5D Lagrangian densities for the various fields, and solve the resulting 5D equations of motion  to obtain the various profiles. Following the strategy outlined in \cite{Albrecht:2009xr}, this can be done before the EWSB. After such a symmetry breaking takes place, one can treat the ratio $v/M_{KK}$ of the Higgs vacuum expectation value (vev) $v$ and the  mass  of the lowest KK mode $M_{KK}$ as a perturbation. The effective 4D Lagrangian is obtained after integration over $y$, and the Feynman rules of the model are worked out neglecting terms of ${\cal O}(v^2/M_{KK}^2)$ or higher. On the same footing, the mixing occurring between  SM fermions and  higher KK fermion modes can be neglected, since  it leads to ${\cal O}(v^2/M_{KK}^2)$ modifications of the relevant couplings. 

As for gauge bosons, we consider KK modes up to the first excitation (1-mode). Indeed, as observed in \cite{buras1}, the model becomes non perturbative already for scales corresponding to the first few KK modes, so that including the whole tower of excitations would lead to unreliable results.

Let us now examine the various sectors of the model, stressing the most relevant features for our analysis.
\subsection{Higgs sector}
After the gauge group (\ref{group}) has undergone the breaking to the SM  $SU(3)_c \times SU(2)_L \times U(1)_Y$,
the electroweak symmetry breaking takes place. A Higgs field $H(x,y)$ is introduced, which transforms as a bidoublet under $SU(2)_L \times SU(2)_R$ and as a singlet under $U(1)_X$. This field contains two charged and  two neutral components:
\be
H(x,y)=\left(\begin{array}{cc}
\frac{\pi^+}{\sqrt{2}} & -\frac{h^0-i\pi^0}{2} \label{Hbidoublet} \\
\frac{h^0+i\pi^0}{2}  & \frac{\pi^-}{\sqrt{2}}
\end{array}\right)  \,\,\, . \nn
\ee
Performing the KK decomposition, one  writes
\be
H(x,y)=\frac{1}{\sqrt{L}}\sum_k H^{(k)}(x) h^{(k)}(y)\,\,.
\label{higgsKK}
\ee
For  the Higgs  localized on the IR brane the choice
\be
h(y)\equiv h^{(0)}(y)  \simeq e^{kL} \delta(y-L)
%h(y)=h^{(0)}(y) = \sqrt{L}\, e^{kL} \delta(y-L)
\label{H-loc}
\ee
is done. As for the components depending on the 4D coordinates, one chooses that
 only the neutral field  $h^0$ has  a non vanishing  vacuum expectation value   coinciding with the SM Higgs vev $v=246.22$ GeV.
The 5D action involving the Higgs field reads:
 \bea
 S_{Higgs}=\hspace*{7cm} \nn \\ \int d^4x \int_0^L dy \sqrt{G} \, {\rm Tr} \Big[ [D_M H(x,y)]^\dagger[D^M H(x,y)] -V(H) \Big]\,\,, \nn \\
 \eea
with $\sqrt{G}=det[g_{MN}]=e^{-4ky}$,  $g_{MN}$ being the 5D metric tensor and $M,N=0,1,2,3,5$.  We do not specify the potential  $V(H)$   since it is irrelevant for our present purposes.
 The covariant derivative $D_M$ involves the gauge bosons of the group (\ref{group}) and is  the starting point to give mass to a number of them.
 Before considering the result of this procedure,  we  define how the various gauge bosons undergo mixing.

\subsection{Gauge boson mixing}
Charged gauge bosons are defined in analogy to  SM,
\be
W_{L(R)\mu}^\pm=\frac{W^1_{L(R)\mu} \mp i W^2_{L(R)\mu}}{\sqrt{2}}\,\,.
\label{chargedW}
\ee
Mixing occurs between the bosons $W_R^3$ and $X$  with a mixing angle $\phi$. The resulting fields are denoted as $Z_{X }$ and $B$:
\bea
Z_{X \mu} &=& c_\phi \, W_{R \, \mu}^3 -s_\phi \, X_\mu \nn \\
B_\mu &=& s_\phi \, W_{R \, \mu}^3 + c_\phi \, X_\mu \,\,\, , \label{1stmix} \eea
where
\bea
c_\phi=\cos{\phi}&=&\frac{g}{\sqrt{g^2+g_X^2}} \nn \\
 s_\phi=\sin{\phi}&=&\frac{g_X}{\sqrt{g^2+g_X^2}}\,\,.
\label{cphi} \eea
In a second step, $W_{L \,}^3$ mixes with $B$ with an angle $\psi$, in complete analogy to   SM, providing the $Z$ and $A$ fields:
\bea
Z_{ \mu} &=& c_\psi \, W_{L \, \mu}^3 -s_\psi \, B_\mu \nn \\
A_\mu &=& s_\psi \, W_{L \, \mu}^3 + c_\psi \, B_\mu \,\,\, ,\label{1stmix} \eea
with
\bea
c_\psi=\cos{\psi}&=&\frac{1}{\sqrt{1+s_\phi^2}} \nn \\
 s_\psi=\sin{\psi}&=&\frac{s_\phi}{\sqrt{1+s_\phi^2}}\,\,.
\label{cpsi} \eea
At the end of the mixing pattern (leaving aside the eight gluons $G_\mu$ with BC $(++)$), we are left with 
\begin{itemize}
\item four charged bosons: $W_L^\pm (++)$ and $W_R^\pm (-+)$;
\item three  neutral bosons: $A(++)$, $Z(++)$ and $Z_{X}(-+)$.
\end{itemize}
We have specified the  BC  for these fields.
For each vector boson field $V_\mu(x,y)$,  the KK expansion is
\be
V_\mu(x,y)=\frac{1}{\sqrt{L}}\sum_{n=0}^{\infty}V_\mu^{(n)}(x) f_V^{(n)}(y) \,\,. \label{KK-vec-bos}
\ee
The free action for each gauge boson  reads
 \be
 S_{gauge}=\int d^5 x \sqrt{G} \left( -\frac{1}{4} F_{MN} F^{MN} \right) \,\,\, ,
 \label{action-gauge}
 \ee
where $F_{MN}$ is the  5D field strength. From (\ref{action-gauge})    the equation of motion for $V_\mu$ can be derived. The solution provides us with the bulk  profiles of each KK mode, $f_V^{(n)}(y)$, which are different if the $(++)$ or $(-+)$ BC are imposed,  but do not depend on the specific  boson.
The  profiles are collected in the Appendix \ref{app:profiles}. Here we only mention that
 \begin{itemize}
\item profiles of zero-modes  are flat,  $f_V^{(0)}(y)=1$;
\item 1-mode profiles for gauge bosons having a zero-mode are denoted by $g(y)$,  and the mass of such modes  is denoted as $M_{++}$;
\item 1-mode profiles for gauge bosons without a zero-mode are denoted by ${\tilde g}(y)$, and  the mass of such modes  is denoted as $M_{-+}$.
\end{itemize}
The  solution of the equation of motion shows that such masses are $M_{++} \simeq 2.45 f$ and $M_{-+} \simeq 2.40 f$, where the dimensionful parameter $f$ is defined as $f=k \, e^{-kL}$.
The numerical value we choose for this parameter is $f=1$ TeV, consistent with other analyses \cite{buras1,buras4}. Hence,
before the EWSB the zero modes of the gauge fields (when present) are massless, while higher KK modes are massive.
The Higgs mechanism occurs  to partially break the symmetry. Since the QCD group $SU(3)$ remains unbroken, as well as  $U(1)_{em}$,   as  in the SM  gluons and  photon do not get mass.
This means that their zero modes remain massless, while higher KK modes are massive but  they do not get a  mass enhancement from the Higgs mechanism. For the remaining fields, mass is acquired and depends on the Higgs vev. Furthermore,
 mixing  among zero modes and higher KK modes occurs. Neglecting modes with KK number larger than $1$, the mixing involves
 \begin{itemize}
\item the charged bosons $W_L^{\pm(0)},\,W_L^{\pm(1)}$ and $W_R^{\pm(1)}$, with the result
\be
\left(\begin{array}{c} W^\pm \\W_H^\pm \\ W^{\prime \pm} \end{array} \right)={\cal G}_W \,\,\left(\begin{array}{c} W_L^{\pm (0)} \\W_L^{\pm (1)} \\ W_R^{ \pm (1)} \end{array} \right) \,\,\, ;
\ee
\item the neutral bosons $Z^{(0)}$, $Z^{(1)}$ and $Z_X^{(1)}$,  giving the mass eigenstates as follows:
\be
\left(\begin{array}{c} Z \\Z_H \\ Z^{\prime} \end{array} \right)={\cal G}_Z \,\,\left(\begin{array}{c} Z^{ (0)} \\Z^{ (1)} \\ Z_X^{  (1)} \end{array} \right) \,\,.
\ee
\end{itemize}
 The expression of the mixing matrices ${\cal G}_W$ and ${\cal G}_Z$, as well as of the  masses
of the mass eigenstates, can be found in Ref.\cite{Albrecht:2009xr}.

\subsection{Fermions}
Fermions are embedded in suitable representations of the gauge group (\ref{group}). We refer  to Ref.\cite{Albrecht:2009xr} for the realization of the fermion sector, and only recall the following issues, holding for  three generations of quarks and leptons, $i=1,2,3$:
\begin{itemize}
\item Left-handed doublets are  in a bidoublet of $SU(2)_L \times SU(2)_R$, together with two new fermions;
\item Right-handed up-type quarks are singlets; no corresponding fields exist in the case of leptons, since the neutrinos are kept left-handed;
    \item Right-handed down-type quarks and charged leptons are  in multiplets that transform as $(3,1) \oplus (1,3)$ under  $SU(2)_L \times SU(2)_R$;  the multiplets contain additional new fermions;
        \item The electric charge reads, in terms of the third component of the $SU(2)_L$ and $SU(2)_R$ isospins and of the charge $Q_X$: $Q=T^3_L+T^3_R+Q_X$.
        \end{itemize}
Since  we  consider only the zero-modes of SM quarks and leptons, we do not  elaborate  on the new fermions.
Solving the equations of motion for ordinary fermions leads to their zero-mode profiles,  denoted as $f_{L,R}^{(0)}(y,c)$ and given in  the Appendix \ref{app:profiles}.  In principle, right and left-handed fermions are treated as distinct fields. The only difference among the fermions resides in the parameter $c$, identified with the fermion mass in the bulk.
$c$ is the same for fields belonging to the same $SU(2)_L \times SU(2)_R$ multiplet: This is the case of $u_L$ and $d_L$, $c_L$ and $s_L$, $t_L$ and $b_L$,  as well as for $\nu_\ell$ and $\ell^-_L$ ($\ell=e,\mu,\tau$).  All the $c$ parameters are chosen real.

An important issue concerns the quark mass eigenstates. As in  SM, they are obtained upon rotation of the flavour eigenstates. We adopt the notation ${\cal U}_{L(R)}$, ${\cal D}_{L(R)}$ for the rotation matrices of the up-type left (right) and down-type left (right) quarks, respectively. The relation $V_{CKM}={\cal U}_L^\dagger {\cal D}_L$ holds.
However, while in  SM the CKM matrix only enters in charged current interactions, here the rotation matrices also modify the neutral currents.
This happens because the integration over the fifth coordinate in the action leads to factors representing overlap integrals of the profiles of two fermions $f_i$ and $f_j$  and a gauge boson profile. These integrals are of two kinds:
\bea
{\cal R}_{f_i f_j} &=& \frac{1}{L} \int_0^L dy  \, e^{ky} \,  f_{f_i}^{(0)}(y,c_i) \, f_{f_j}^{(0)}(y,c_j) \, g(y) \nn \\
\tilde{\cal R}_{f_i f_j} &=& \frac{1}{L} \int_0^L dy  \,e^{ky} \,  f_{f_i}^{(0)}(y,c_i) \, f_{f_j}^{(0)}(y,c_j) \, {\tilde g}(y) \,\,.
 \eea
Before EWSB the interaction is flavour diagonal, so that the overlap integrals can be collected in two  matrices
${\cal R}_f=diag \left({\cal R}_{f_1 f_1}, {\cal R}_{f_2 f_2}, {\cal R}_{f_3 f_3}\right)$ and ${ \tilde{\cal R}}_f=diag \left(\tilde{\cal R}_{f_1 f_1}, \tilde{\cal R}_{f_2 f_2}, \tilde{\cal R}_{f_3 f_3}\right)$. After the rotation to mass eigenstates, one is left with a typical product  ${\cal M}^\dagger {\cal R}_f {\cal M}$, where ${\cal M}={\cal U}_{L,R}, {\cal D}_{L,R}$, so that one can no more exploit the relation:  ${\cal M}^\dagger {\cal M}=1$ and FCNC are induced already at tree level. They are mediated by the threee neutral EW gauge bosons $Z,\,Z^\prime, \, Z_H$ as well as by the first KK mode of the photon and of the gluon, although the latter does not contribute to processes with leptons in the final state.
 For this reason, the quark rotation matrices appear in the Feynman rules of this theory: we collect in the Appendix \ref{app:feynman} a  few rules needed in our analysis.

The expression of the elements of the rotation matrices are required. We refer to \cite{buras1} for the list of the various entries. Here we only mention that they all are written in terms of the quark profiles and of the 5D Yukawa couplings  which are denoted by $\lambda_{ij}^u$ for up-type quarks and $\lambda_{ij}^d$ for down type quarks, respectively.
On the other hand, the effective 4D Yukawa couplings can be defined as:
\be
Y_{ij}^{u(d)}=\frac{1}{\sqrt{2}}\frac{1}{L^{3/2}} \int_0^L \, dy \, \lambda_{ij}^{u(d)} f_{q_L^i}^{(0)}(y)f_{u_R^j(d_R^j)}^{(0)}(y) h(y)
\label{Yud}
\,\,. 
\ee
Since the fermion profiles depend exponentially on the bulk mass parameters (see Appendix \ref{app:profiles}), one identifies in the above relation the origin of the hierarchy of fermion masses and mixing \cite{gherghetta,grossman-neubert}.

As for the matrices ${\cal U}_{L(R)}$ and ${\cal D}_{L(R)}$,
 not all their elements are independent, since the Yukawa couplings determine the quark masses and since the product $V_{CKM}={\cal U}_L^\dagger {\cal D}_L$ should be satisfied, as already mentioned.
In particular, the relations hold:
\bea
m_u&=& \frac{v}{\sqrt{2}}\frac{det(\lambda^u)}{\lambda^u_{33}\lambda^u_{22}-\lambda^u_{23}\lambda^u_{32}}\frac{e^{kL}}{L}f_{u_L}f_{u_R} \nn \\
m_c &=& \frac{v}{\sqrt{2}} \frac{\lambda^u_{33}\lambda^u_{22}-\lambda^u_{23}\lambda^u_{32}}{\lambda^u_{33}}\frac{e^{kL}}{L}f_{c_L}f_{c_R} \label{u-masses} \\
m_t&=&\frac{v}{\sqrt{2}}\lambda^u_{33}\frac{e^{kL}}{L}f_{t_L}f_{t_R}\,\,\, , \nn \eea
as well as the analogous relations for down-type quarks with the substitution $\lambda^u \to \lambda^d$. 
We have adopted the short notation: $f_{q_{L,R}}=f_{q_{L,R}}^{(0)}(y=L,c_{q_{L,R}})$.

To understand how many among the remaining entries should be considered as independent ones, we adopt further simplifications. In particular, the entries of the matrices $\lambda^{u,d}$ are treated as real numbers, since the effects that we are interested to investigate involve CP conserving observables, hence they do not require the introduction of new phases besides those present in SM. Therefore,  after imposing the quark mass constraints, we are left with  six independent entries among the elements of the Yukawa matrices, that we choose to be\footnote{A  parametrization of the matrices $\lambda^{u,d}$ that considers complex entries can be found in \cite{buras1}.}
\bea
\lambda^u_{12} \,\,\, , \hskip 0.5 cm  \lambda^u_{13} \,\,\, ,\hskip 0.5 cm \lambda^u_{23} \,\,\, ,\nn \\
\lambda^d_{12} \,\,\, ,\hskip 0.5 cm  \lambda^d_{13} \,\,\, ,\hskip 0.5 cm \lambda^d_{23}  \,\,\, .
\label{lambdas}
\eea
Together with the bulk mass parameters, these constitute the set of numerical input in our study. The way we treat them is described in the Section \ref{sec:num}.

\section{Modification of the Wilson coefficients in RS$_c$ model}
In the RS model the Wilson coefficients in the effective Hamiltonian (\ref{hamil}) are modified with respect to  SM:
\be
C_i^{(\prime)}=C_i^{(\prime)SM}+\Delta C_i^{(\prime)} \,\,,  \,\,\,\,i=7,\,9,\,10 \,\,\, .\label{RS-coeff} \ee
We neglect the tiny SM contribution to the primed coefficients, when present,  while for the unprimed coefficients 
%at the scale $\mu_b \simeq {\cal O}(m_b)$ 
we use:
\bea
C_7^{SM}(\mu_b) &=& -0.301\nn \\
C_9^{SM} (\mu_b) &=& 4.07 \label{SMcoef-num} \\
C_{10}^{SM} &=&-4.31 \nn \,\,,\eea
where $\mu_b =2.5$ GeV.
In the  RS$_c$ model  the results  for $\Delta C_{9,10}^{(\prime)}$, derived in \cite{buras2} at the high scale $\mu=M_{KK}$, read:
\bea
\Delta C_9&=&\left[\frac{\Delta Y_s}{\sin^2(\theta_W)}-4 \Delta Z_s \right] \,\, , \nn \\
\Delta C_9^\prime&=&\left[\frac{\Delta Y_s^\prime}{\sin^2(\theta_W)}-4 \Delta Z_s^\prime \right] \,\, , 
\nn \\
\Delta C_{10}&=&- \frac{\Delta Y_s}{\sin^2(\theta_W)} \,\, ,  \\
\Delta C_{10}^\prime&=&-\frac{\Delta Y_s^\prime}{\sin^2(\theta_W)}\,\, , \nn
 \eea
where
\bea
\Delta Y_s&=&-\frac{1}{V_{tb}V_{ts}^*} \sum_{X} \frac{\Delta_L^{\ell \ell}(X) -\Delta_R^{\ell \ell}(X)}{4 M_X^2 g_{SM}^2}\Delta_L^{bs}(X) \,\, , \nn \\
\Delta Y_s^\prime&=&-\frac{1}{V_{tb}V_{ts}^*} \sum_{X} \frac{\Delta_L^{\ell \ell}(X) -\Delta_R^{\ell \ell}(X)}{4 M_X^2 g_{SM}^2}\Delta_R^{bs}(X) \,\, , \nn \\
\Delta Z_s&=&\frac{1}{V_{tb}V_{ts}^*} \sum_{X} \frac{\Delta_R^{\ell \ell}(X)}{8 M_X^2 g_{SM}^2 \sin^2(\theta_W)}\Delta_L^{bs}(X) \,\, , \\
\Delta Z_s^\prime&=&\frac{1}{V_{tb}V_{ts}^*} \sum_{X} \frac{\Delta_R^{\ell \ell}(X)}{8 M_X^2 g_{SM}^2 \sin^2(\theta_W)}\Delta_R^{bs}(X) \,\,\, .\nn \eea
The sums run over the neutral bosons $X=Z, \, Z_H,\,Z^\prime$ and $A^{(1)}$, with  $\dd g_{SM}^2=\frac{G_F}{\sqrt{2}}\frac{\alpha}{2 \pi \sin^2(\theta_W)}$, and $\theta_W$  the Weinberg angle.
The functions $\Delta_{L,R}^{f_i f_j}(X)$,  encoding the couplings of the $X$ bosons to the fermions $f_i,f_j$,  are collected in the Appendix \ref{app:feynman}.
$\Delta C_{9,10}^{(\prime)}$ 
%are renormalization group invariant, and 
do not need to be evolved to  $\mu_b$.

The case of $\Delta C_7^{(\prime)}$ is different. 
In Ref.~\cite{Blanke:2012tv} a determination of this coefficient in the RS model with and without custodial protection has been carried out directly in 5D, using the mixed position/momentum formalism. This approach includes  the contribution of  the whole tower of KK excitations. However,  since the other coefficients used  in this paper  have been computed  in  the effective 4D model, we have repeated  the calculation for $C_7(M_{KK})$ in 4D, as described in Appendix \ref{app:c7}, with a  set of assumptions concerning the contributions of the KK excitations consistent with the calculation of $\Delta C_{9,10}^{(\prime)}$.
 In particular, we have kept only the dominant contribution of the first KK mode in the case of the intermediate gluon and Higgs fields exchanged in the diagrams in Fig.~\ref{fig:diagC7}.  For the intermediate fermions, consistently with the procedure described in the previous Section we have included only the zero modes. 
For the evolution at the scale $\mu_b$,  we use the master formula   \cite{Blanke:2012tv}:
\be
\Delta C_7^{(\prime)}(\mu_b)=0.429 \, \Delta C_7^{(\prime)}(M_{KK})+0.128 \, \Delta C_8^{(\prime)}(M_{KK}) \,\,
\ee
which shows that   $\Delta C_8^{(\prime)}(M_{KK}) $ are needed; they are also computed in Appendix \ref{app:c7}.

\section{Numerical analysis}\label{sec:num}
The results for the  observables  considered in this paper within the RS$_c$ model are obtained adding the new contributions to the Wilson coefficients  computed  scanning  the  parameter space of the model. 
In particular, we focus on  the elements of the two Yukawa matrices $\lambda^{d,u}$, and  on the bulk mass parameters for quarks and leptons.

 The diagonal elements of  $\lambda^{d,u}$ are fixed from the  relations (\ref{u-masses}) (and the analogous ones for down-type quarks),  so that, under the assumptions described in Sec. \ref{sec:RS}, we scan over the two sets in Eq.(\ref{lambdas}). As customary for these scenarios, the requirement of perturbativity of the model up to the scale of the first three KK modes sets the range: $|\lambda^{d,u}_{ij}| \le 3/k$. 
However, not all the values in this range are acceptable,  since also the CKM matrix elements should be reproduced after  the constraint  $V_{CKM}={\cal U}_L^\dagger {\cal D}_L$ is imposed. 
The first step in the parameter selection  consists in fixing the bulk mass terms $c_i$ for down- and up-type quarks. 
Several analyses have been devoted to this purpose in the literature. We adopt the strategy outlined in \cite{buras1}  and the  consequent choice of parameters   \cite{Duling:2010lqa}. 
It consists in imposing that quark mass parameters  at the high scale ${\cal O}(M_{KK})$, obtained from the $\overline{MS}$ masses using NLO  renormalization group evolution,  and CKM elements are reproduced within $2\sigma$. An exception  is represented by the bulk mass parameter of the left-handed doublet of the third generation of quarks. We slightly vary it in a range which  also satisfies the constraints  derived in \cite{neubertRS} exploiting  the experimental measurements of several quantities related to $Z$ decays to $b$ quarks, i.e. the coupling $Z{\bar b}b$, the $b$-quark left-right asymmetry parameter and the forward-backward asymmetry for $b$ quarks  \cite{ALEPH:2005ab}. 
The set   used in our analysis  is
\footnote{The fermion profile given in Appendix \ref{app:profiles} corresponds to the case of a left-handed fermion with bulk mass parameter $c_L$.
For right-handed fermions one should use the same function reversing the sign of the bulk mass parameter $c_R$.
Since $c_L$ and $c_R$ are independent of each other and vary in the range $[-1,1]$, we can choose to adopt the same profile for both fermions. However, we  reverse the sign of the numerical solution found in \cite{Duling:2010lqa} for the parameters $c_R$.}:
\bea
c_L^{u,d}=0.63 \,\, , \hskip 0.3 cm   c_L^{c,s}=0.57 \,\, ,\hskip 0.3 cm c_L^{b,t}\in [0.40,\, 0.45] \,\, , \,\,\, \nn \\
c_R^u = 0.67 \,\, ,\hskip 0.3 cm c_R^c=0.53 \,\, ,\hskip 0.3 cm c_R^t=-0.35 \,\, , \,\,\,\,\, \label{bulkmasses}  \\
c_R^d = 0.66 \,\, , \hskip 0.3 cm c_R^s=0.60 \,\, ,\hskip 0.3 cm c_R^b=0.57 \,\,\,\,. \,\,\,\,\,\,\,\,\, \nn
\eea

For  leptons, $c_\ell$ are set to $c_\ell=0.7$  in all cases, motivated by the observation  that lepton flavour-conserving couplings are almost independent of the choice of their bulk mass parameter  provided that $c_\ell>0.5$  \cite{buras2}. Other determinations  can be found in \cite{Huber:2000ie,burdman,agashe,Agashe:2004ay,Fitzpatrick:2007sa,neubertRS,Archer:2011bk}.

Fixed such values, we generate the six $\lambda$ parameters in (\ref{lambdas}) which also satisfy  the CKM constraints.
%as  specified in \cite{buras1,Duling:2010lqa}. 
In particular, we impose  $|V_{cb}|$ and $|V_{ub}|$ in   the largest range found from their experimental determinations from inclusive and exclusive $B$ decays \cite{Amhis:2012bh}, and impose that $|V_{us}|$ lies within $2\%$ of the central value quoted by PDG \cite{Beringer:1900zz}:
\bea
|V_{cb}| &\in& [0.038,\,0.043] , \nn \\
 |V_{ub} |&\in& [0.00294,\,0.00434] ,   \\
| V_{us} |&\in& [0.22,\,0.23] . \nn
\label{VxbVus}
\eea
For the quark masses we use
\be
m_d= 4.9 \,\, {\rm MeV} , \hskip 0.1 cm m_s=90 \,\, {\rm MeV}  ,  \hskip 0.1  cm  m_b= 4.8 \,\, {\rm GeV}  \,\,\, .
\ee
\begin{figure}[t!]
\includegraphics[width = 0.4\textwidth]{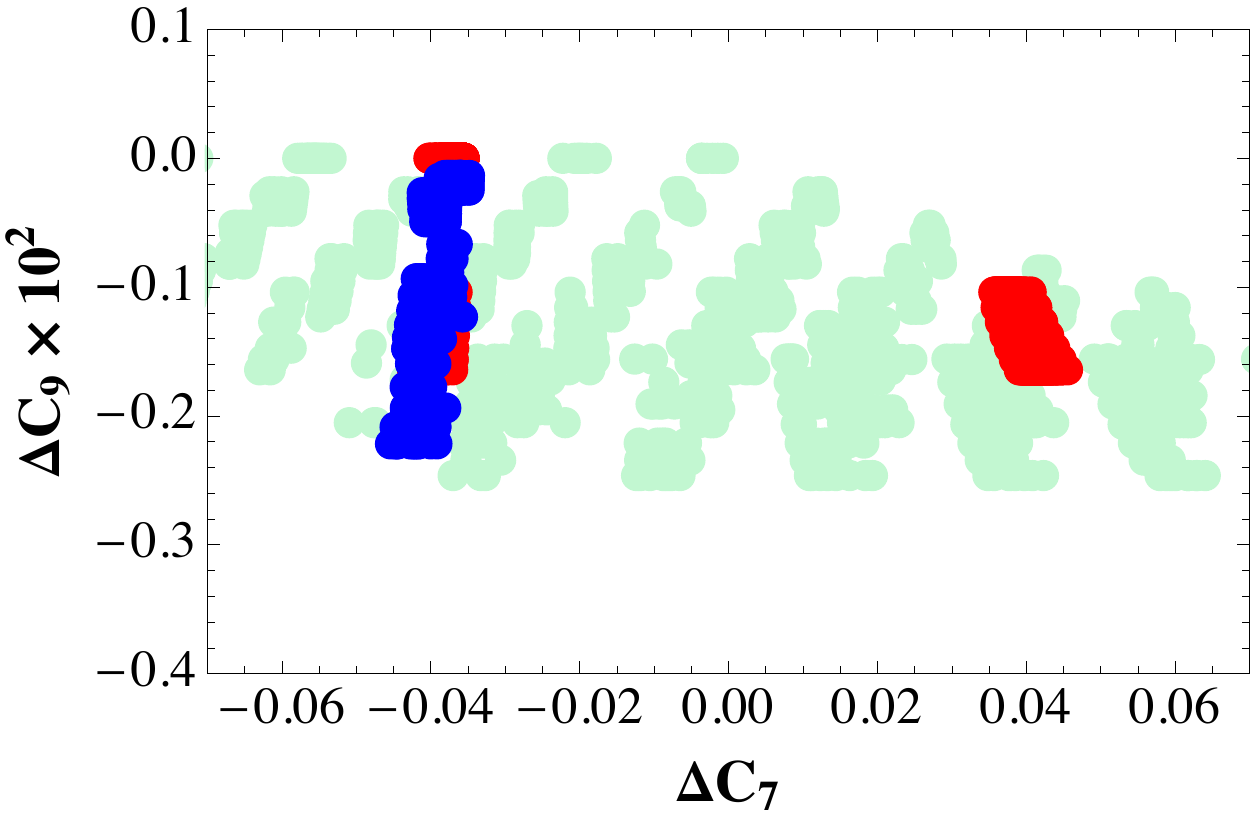}
\caption{ $\Delta C_7(m_b)$ vs $\Delta C_9$  obtained  implementing sequentially the constraints described in the text. The  light  green points correspond to the constraints from $|V_{us}|$ and $|V_{ub}|$, the red  points to the  constraint from $|V_{cb}|$ and $|V_{ub}|$,
the blue  points to the further constraints from ${\cal B}(B \to K^* \mu^+ \mu^-)_{exp}$ and ${\cal B}(B \to X_s \gamma)_{exp}$.   }\label{fig:deltac7s}
\end{figure}
These  constraints are  the starting point  of our analysis. The generated values of the parameters  fulfilling all the constraints are used to compute 
 the RS contributions to the Wilson coefficients.   Two further  conditions on  the  computed $B \to K^* \mu^+ \mu^-$ and $B \to X_s \gamma$ branching fractions are imposed,  requiring that they are less than
$2\sigma$ from   the experimental measurements 
\bea
\hskip -0.2cm {\cal B}(B \to K^* \mu^+ \mu^-)_{exp} &=& (1.02 \pm^{0.14}_{0.13} \pm0.05)\times 10^{-6} \, ,\,\,\,\,\,  \label{Kstarmumuexp} \\
\hskip -0.2cm{\cal B}(B \to X_s \gamma)_{exp} &=& (3.43 \pm 0.21 \pm 0.07) \times 10^{-4} ;\,\,\,\,\,\,\,\, \label{BXsgammaexp} 
\eea
the result in (\ref{Kstarmumuexp}) is the average performed by BaBar Collaboration of the branching fractions of the four modes $B^{+,0} \to K^{* +,0} \mu^+ \mu^- (e^+ e^-)$  \cite{Lees:2012tva}, while the result in (\ref{BXsgammaexp}) is the HFAG Collaboration average   \cite{Amhis:2012bh}.
An example of the sequence of the effects of the constraints is shown in Fig.\ref{fig:deltac7s}. After imposing the constraints on $V_{ub}$ and $V_{us}$, a set of  values of
$\Delta C_7(m_b)$ and $\Delta C_9$ is computed, and $\Delta C_7(m_b)$ spans a quite broad range of positive and negative values, the green region in the figure. Implementing the constraint on $V_{cb}$ reduces the possibilities to two isolated regions, the red spots in the figure, and the region of negative values, the blue one,  survives after the constraints (\ref{Kstarmumuexp}) and (\ref{BXsgammaexp}) are imposed.
We have checked that the selected points reproduce also the other CKM elements within their uncertainty, except for $|V_{td}|$ which lies in the  $3\sigma$  range around its central value  \cite{Beringer:1900zz}.
 Moreover, as discussed in \cite{buras1}, the set of parameters selected imposing the quark masses and CKM  constraints allows to satisfy also the constraint from ${\bar B}_s - B_s$ mixing. On the basis of this we have not repeated such an analysis, considering that we also impose that the data in (\ref{Kstarmumuexp}-\ref{BXsgammaexp}) are reproduced, which usually represent more severe conditions.

\begin{figure*}[t!]
\begin{tabular}{cc}
\includegraphics[width = 0.4\textwidth]{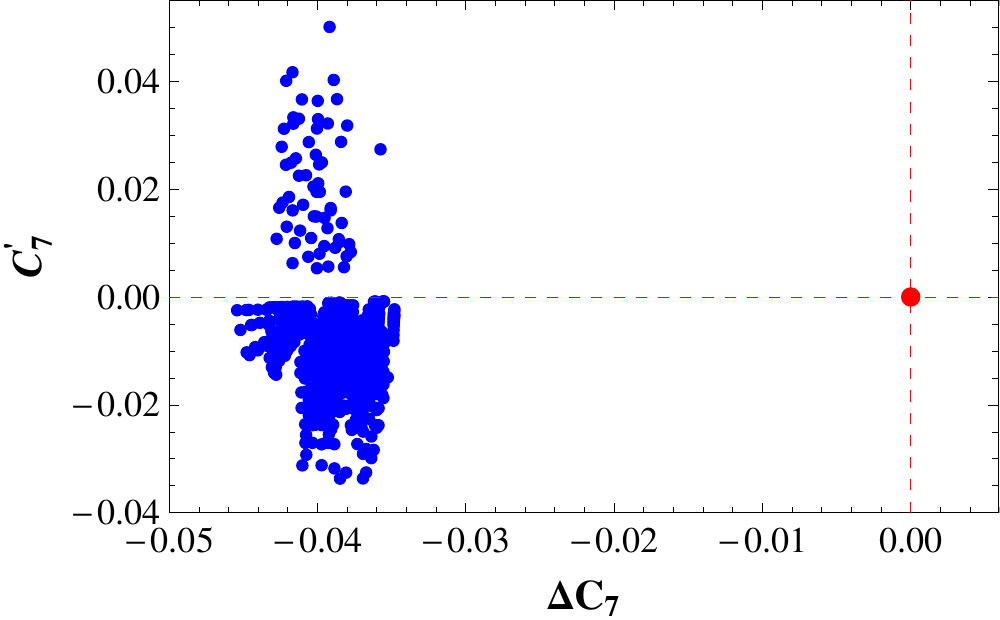} \hspace*{0.4cm} & \hspace*{0.3cm} \includegraphics[width = 0.4\textwidth]{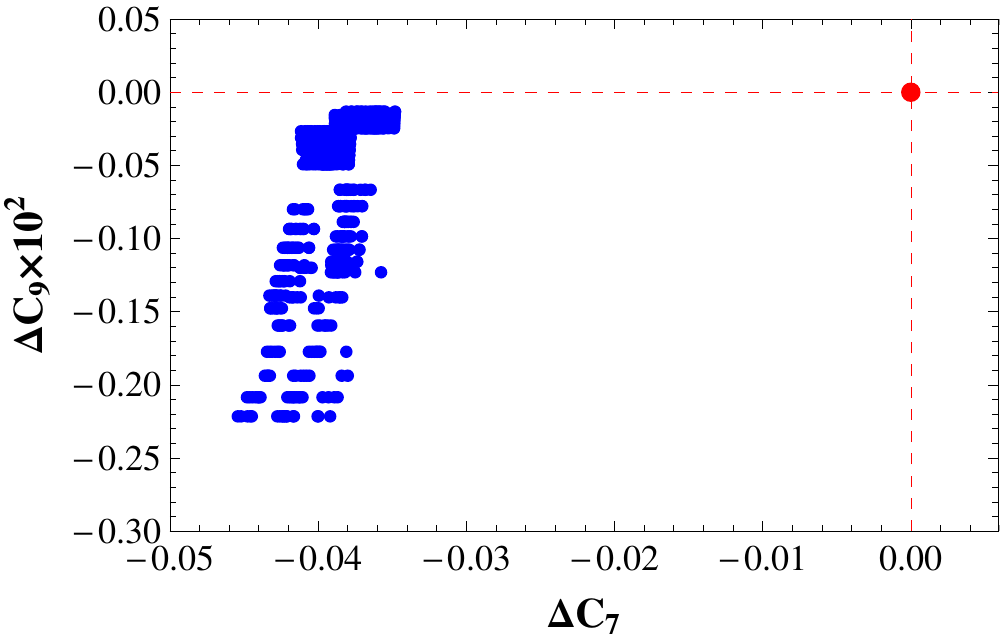}\\ \hspace*{0.5cm}  (a) & \hspace*{1.2cm} (b)
\\ 
\includegraphics[width = 0.4\textwidth]{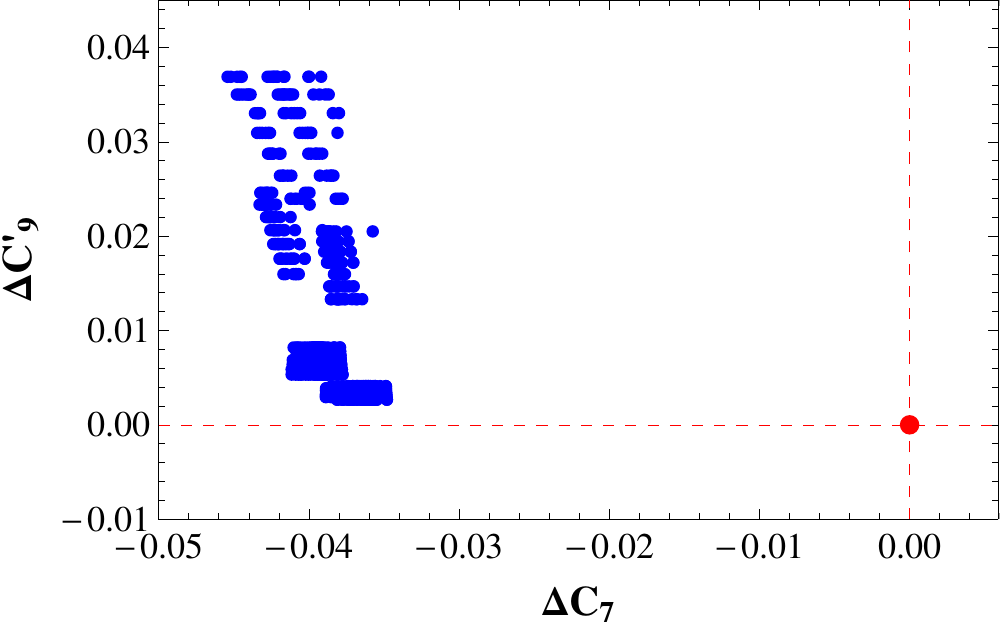}  \hspace*{0.4cm}&\hspace*{0.3cm} \includegraphics[width = 0.4\textwidth]{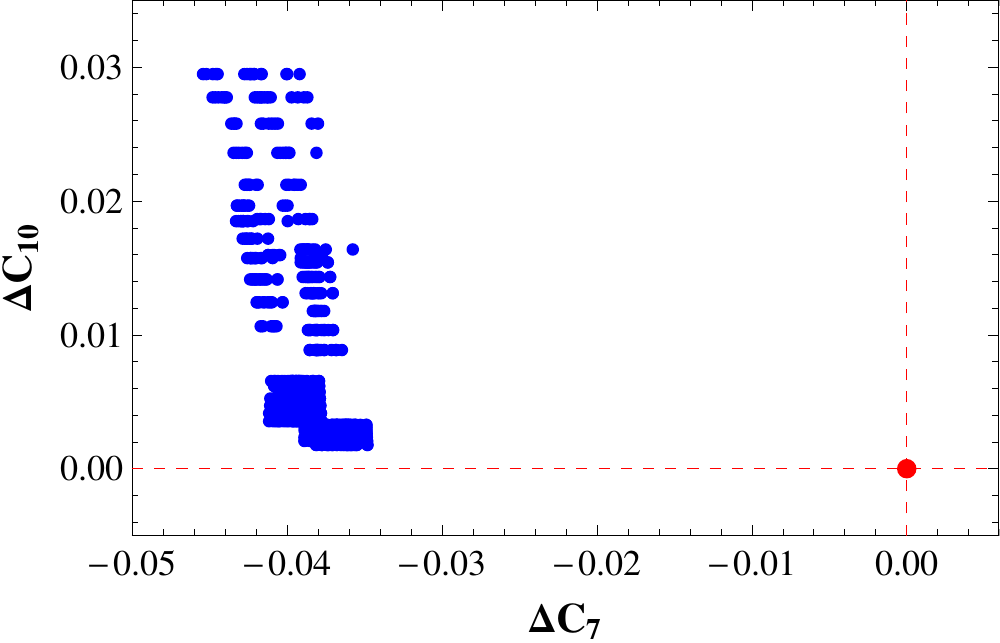} \\ \hspace*{0.5cm}(c) & \hspace*{1.2cm} (d)
\\
\includegraphics[width = 0.4\textwidth]{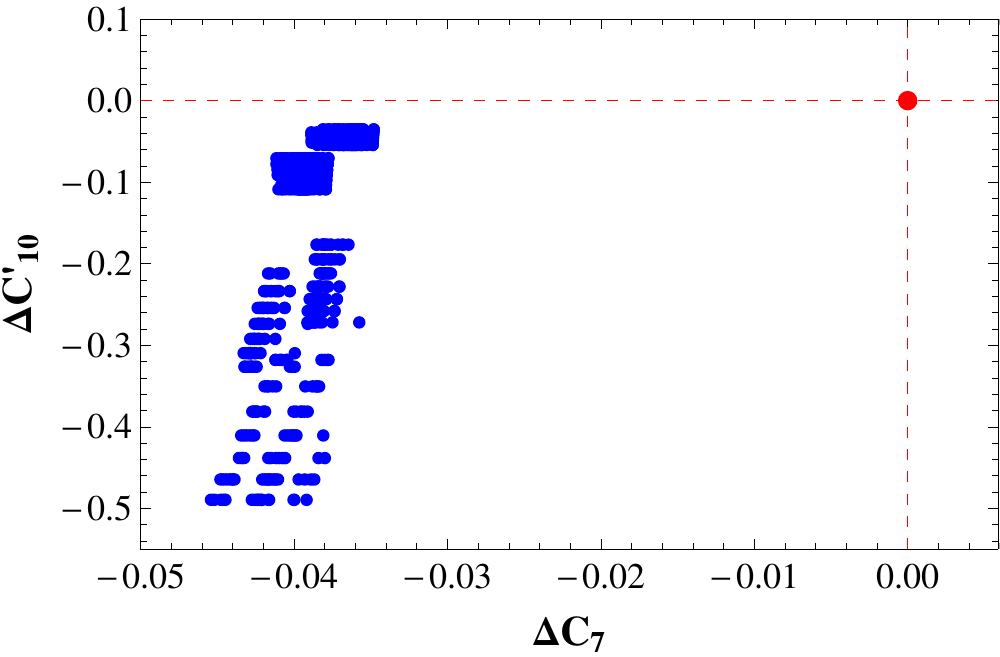}  \hspace*{0.4cm}&\hspace*{0.3cm} \includegraphics[width = 0.4\textwidth]{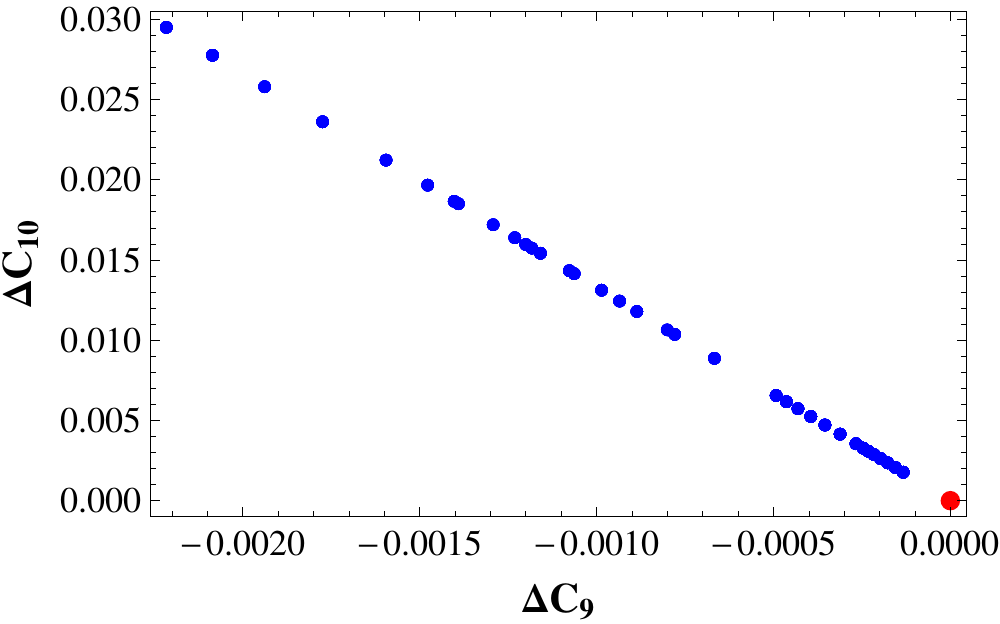} \\ \hspace*{0.5cm}(e) & \hspace*{1.2cm} (f)
\end{tabular}
\caption{Correlations between the RS$_c$ contribution to the Wilson coefficients $C_{7,9,10}^{(\prime)}$. The coefficients  $\Delta C_7^{(\prime)}$  are evaluated at the scale $\mu_b=m_b$. No correction corresponds to the red dot.}\label{fig:corr1}
\end{figure*}

We depict in Fig.\ref{fig:corr1} the obtained values and correlations for  $\Delta C_i^{(\prime)}$. The largest deviations from the SM are 
$|\Delta C_7|_{max}\simeq 0.046\,$, $|\Delta C_7^\prime|_{max}\simeq 0.05\,$, $|\Delta C_9|_{max}\simeq 0.0023\,$, $|\Delta C_9^\prime|_{max}\simeq 0.038\,$, $|\Delta C_{10}|_{max}\simeq 0.030\,$, $|\Delta C_{10}^\prime|_{max}\simeq 0.50\,$. As shown in  the  panel $(f)$ of  the figure, $\Delta C_9$ and $\Delta C_{10}$ are linearly correlated, and the same happens for each pair  $\Delta C_i^{(\prime)}$, $i=9,10$. Indeed,   in the large set of parameters  the most relevant input for these coefficients is $\lambda_{23}^d$,  and  the relations  approximately  hold (for $c_L^{b,t}$ fixed to the central value):
\bea
\Delta C_9 &\simeq& -7.18 \, 10^{-4} \, \lambda_{23}^d \, k \nn \\
\Delta C_9^\prime &\simeq& 1.22  \, 10^{-2} \, \lambda_{23}^d \, k\nn \\
\Delta C_{10} &\simeq& 9.55 \, 10^{-3} \, \lambda_{23}^d \, k  \\
\Delta C_{10} ^\prime &\simeq& -1.62  \, 10^{-1} \, \lambda_{23}^d \, k\nn \,\,.\eea

There have been  attempts to understand what would be the required size of deviations from SM values for the Wilson coefficients that could explain the observed anomalies in $B \to K^* \mu^+ \mu^-$ distributions (on which we shall elaborate below)
and, consequently, which NP scenario might provide such deviations \cite{Descotes-Genon:2013wba,Altmannshofer:2013foa,Beaujean:2013soa}. Although there is not a unique answer to this question, a possible conclusion is that NP  should provide a large negative value of $\Delta C_9$. For example, in \cite{Altmannshofer:2013foa}  various possibilities are considered in which NP affects just one Wilson coefficient, a pair of them, or all of them simultaneously. In the last case, to which the RS scenario belongs, it is found that  the required deviation of $C_9$ from its SM value should be $\Delta C_9 \simeq -0.9$ for a real coefficient, or even $|\Delta C_9 |  \simeq 2.25$  for a complex one. The result of  our analysis shows that such a huge deviation is not  reached in RS$_c$, as  in the NP models considered so far for this purpose. The largest deviations  (still not sizable enough) are found in models introducing a new neutral gauge boson $Z^\prime$ with suitable FCNC couplings to quarks  \cite{Buras:2013qja,Buras:2013dea,Gauld:2013qba}.  

\section{$B^0 \to K^{*0} \ell^+ \ell^-$ observables}\label{sec:results}
\begin{figure}[h]
\includegraphics[width = 0.4\textwidth]{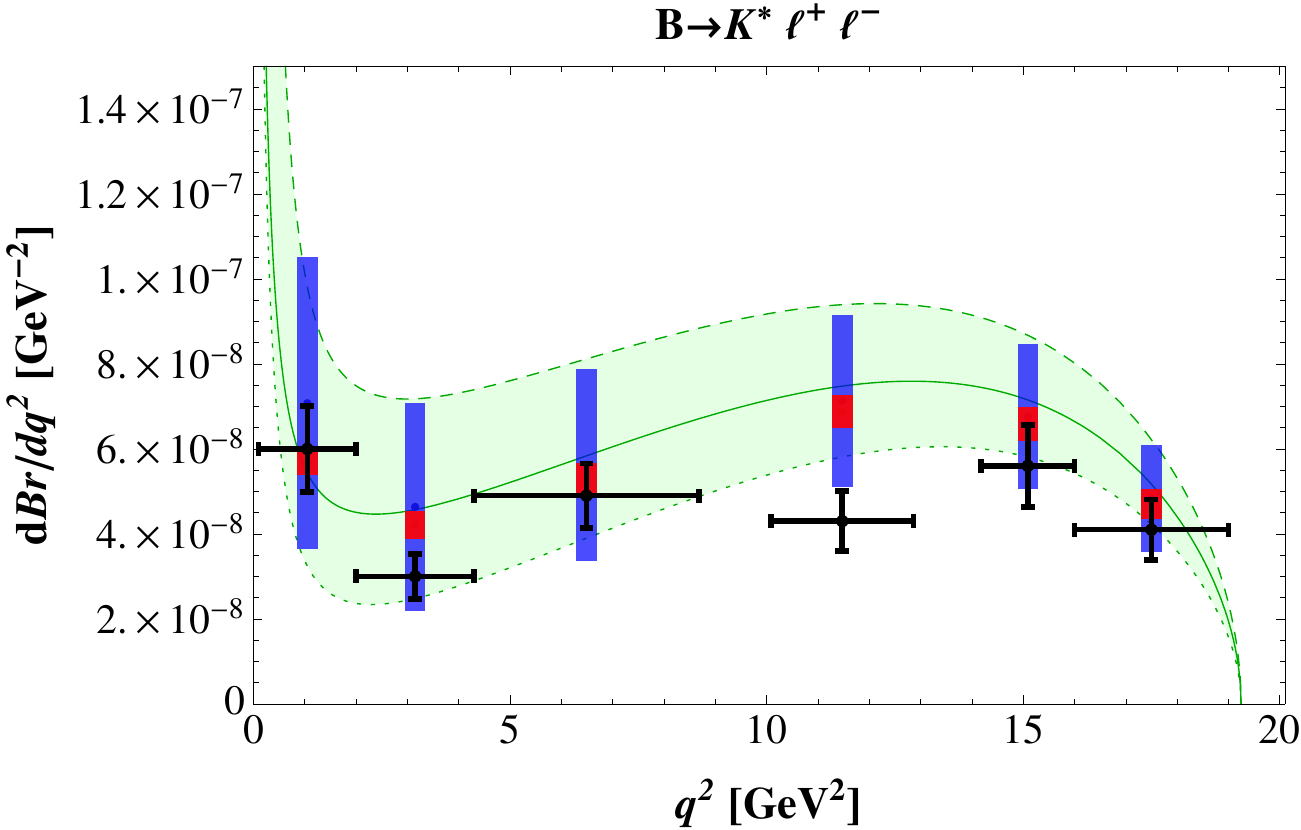}
\caption{Differential $B^0\to K^{*0} \mu^+ \mu^-$ decay rate. The light (green) band corresponds to the SM result, including the uncertainty of  the form factors. The red and blue vertical bars correspond to the RS$_c$ result, 
without or with  the uncertainty in form factors. The black dots, with their error bars, are the LHCb measurements in  \cite{Aaij:2013iag}.}\label{fig:spectr}
\end{figure}

\begin{figure}[h]
\includegraphics[width = 0.4\textwidth]{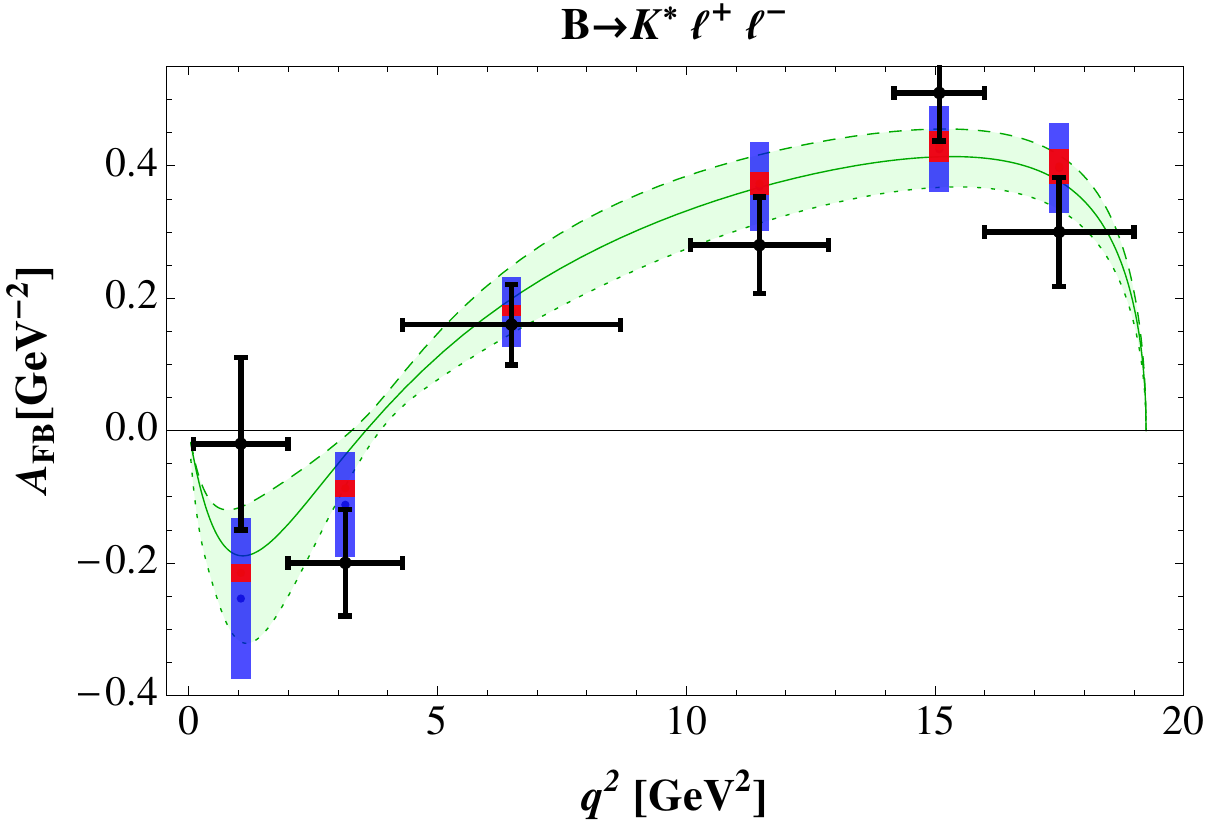}
\caption{Lepton FB asymmetry in  $B^0\to K^{*0} \mu^+ \mu^-$. The symbols have the same meaning  as in Fig.~\ref{fig:spectr}.}\label{fig:AFB}
\end{figure}

\begin{figure}[h]
\includegraphics[width = 0.4\textwidth]{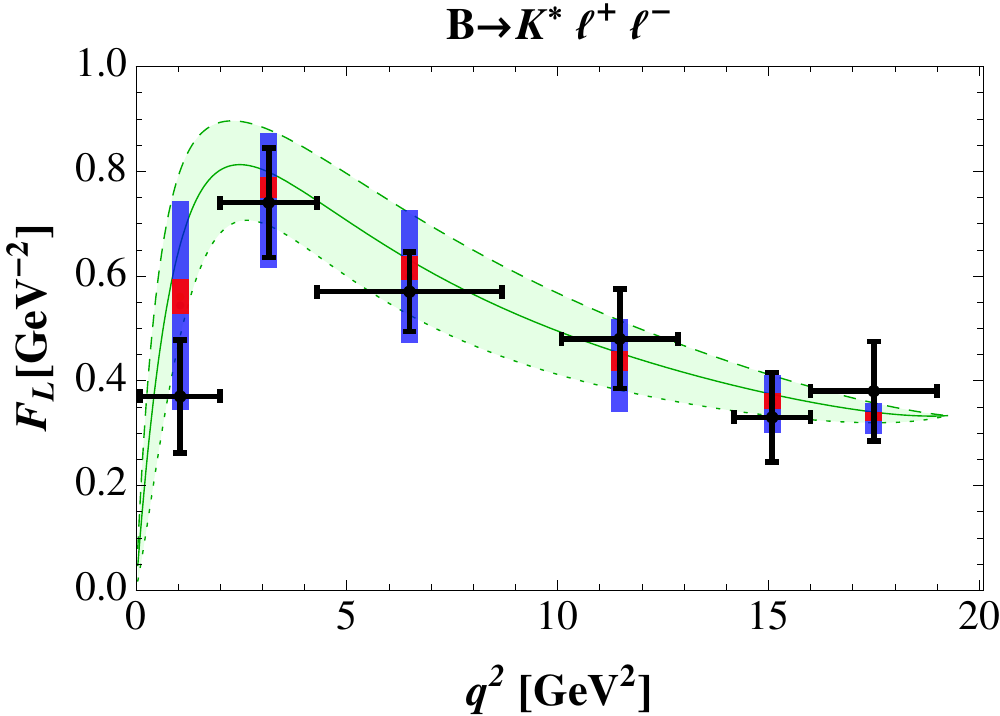}
\caption{  $K^*$ longitudinal polarization fraction in $B^0\to K^{*0} \mu^+ \mu^-$. The symbols have  the same meaning  as in Fig.~\ref{fig:spectr}.}\label{fig:FL}
\end{figure}

\begin{figure}[h!]
\includegraphics[width = 0.4\textwidth]{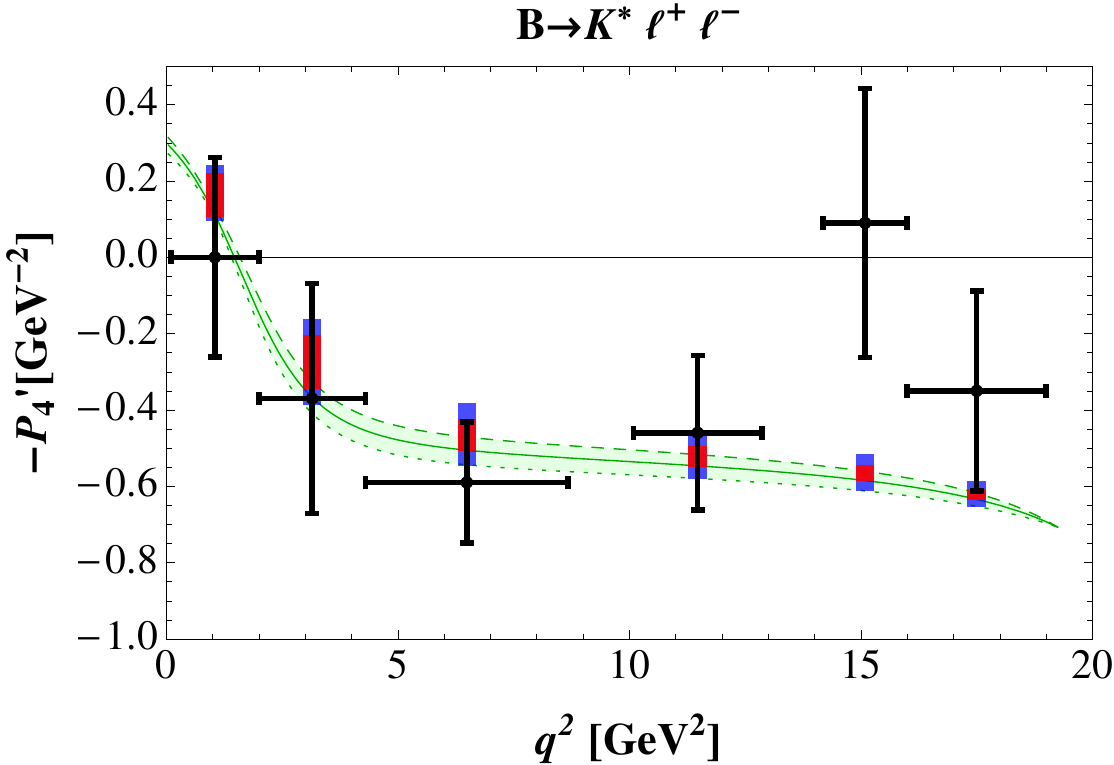}
\caption{Observable $P^\prime_4$ in  $B^0\to K^{*0} \mu^+ \mu^-$.  The light (green) band corresponds to the SM result, including the uncertainty of  the form factors. The red and blue vertical bars correspond to the RS$_c$ result, 
without or with  the uncertainty in form factors. The black dots, with their error bars, are the LHCb measurements in Ref.~\cite{Aaij:2013qta}. The sign is fixed to make   the definition (\ref{Piprime}) and the one  in Ref.~\cite{Aaij:2013qta} compatible. }\label{fig:P4p}
\end{figure}

\begin{figure}[h!]
\includegraphics[width = 0.4\textwidth]{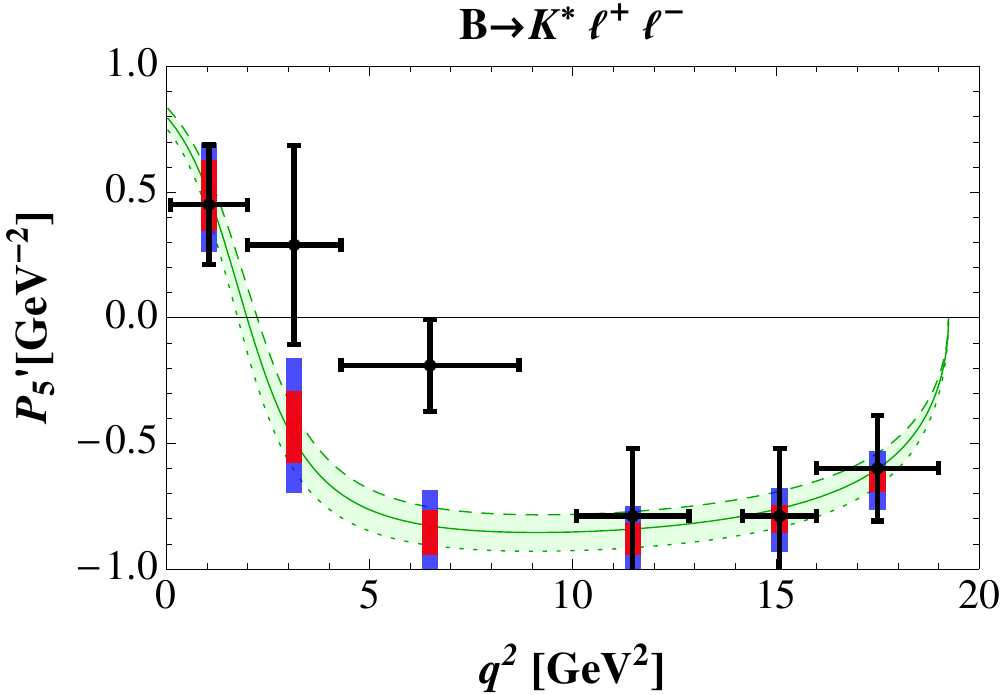}
\caption{Observable $P^\prime_5$ in  $B^0\to K^{*0} \mu^+ \mu^-$.  The meaning of the symbols  is the same as in Fig.~\ref{fig:P4p}. }\label{fig:P5p}
\end{figure}

\begin{figure}[h]
\includegraphics[width = 0.4\textwidth]{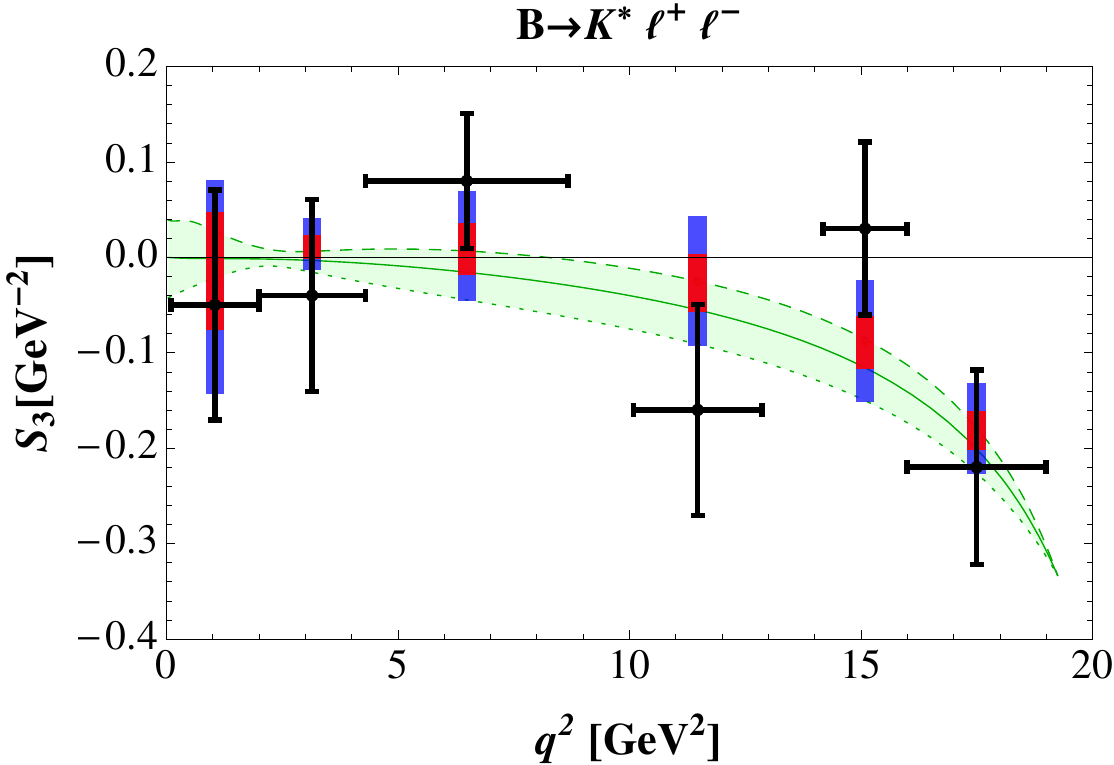}
\caption{ Observable $S_3$ in  $B^0\to K^{*0} \mu^+ \mu^-$. The symbols have  the same meaning  as in Fig.~\ref{fig:spectr}.}\label{fig:S3}
\end{figure}

It is now possible to compute the set of $B \to K^* \mu^+ \mu^-$ observables measured by LHCb, and compare the outcome with data. The results are collected in Figs.\ref{fig:spectr}-\ref{fig:S3}. The  results obtained in SM include the uncertainty in the hadronic form factors; for such nonperturbative quantities we use the light-cone QCD sum rule determination in
\cite{Ball:2004rg} (previous determinations, such as the one in \cite{Colangelo:1995jv}, have larger uncertainties).  The hadronic errors have an impact mainly on  the  $B^0 \to K^{*0} \mu^+ \mu^-$  differential decay rate and on the  $K^*$ longitudinal polarization distribution, while to position of the zero in $A_{FB}(q^2)$ and of the maximum of $F_L$ are less affected, as expected. 

In the analysis of the modifications in RS$_c$, for the various observables, we separately consider  the changes due to the new Wilson coefficients, and the changes which also include 
the hadronic form factor  uncertainties. The results show that the deviations induced in RS$_c$ are small, since the  corrections $\Delta C_{9,10}$ are tiny fractions of 
$C_{9,10}^{SM}$ and that also the coefficients of operators absent in SM, $\Delta C^\prime_{9,10}$ are small. A little effect is found at small $q^2$, where the changes due to 
$\Delta C_{7}^{(\prime)}$  are slightly larger. The comparison of data with predictions confirmes the agreement, excluding the measurement of $A_{FB}(q^2)$ in the first bin of $q^2$ where the predictions are 
larger than the experimental result.   In the high $q^2$ range  the hadronic uncertainty for the lepton FB asymmetry is about $20 \%$.
 
The results for the  observables  $P^\prime_4$,  $P^\prime_5$ and  $S_3$ are  shown in Figs.~\ref{fig:P4p}, \ref{fig:P5p} and \ref{fig:S3}.  In $P^\prime_5$ the hadronic uncertainty is at the level of $10\%$ in all the $q^2$ range. The modification of the prediction obtained in RS$_c$ is similar or larger at low $q^2$, up to $q^2 \simeq 7$ GeV$^2$, therefore this is a favourable kinematical range where to investigate this observable. 
The discrepancy with the measurement in the third $q^2$ bin still persists, while there is agreement  in the other bins.

The hadronic uncertainties turn out to be  smaller in $P^\prime_4$, and the changes in the predictions in RS$_c$ seem promising to be observed at low $q^2$. A deviation observed in the fifth $q^2$ bin of the series of measurements is at the level of less than $2\sigma$.

At odds with $P^\prime_5$, in the observable $S_3$ the RS$_c$ result is systematically above the SM for the largest part of the parameter space, in particular in the large $q^2$ range, as shown in  Fig.~\ref{fig:S3}.   The size of such effect  is comparable with  the hadronic uncertainty; 
therefore, one can envisage the possibility of using this observable for a better characterization of the deviations obtained in this new physics scenario. The experimental results follow the predictions,  but   the errors are too large to draw conclusions.
All the observations can be quantified as done in Table~\ref{tab:chi2}, which confirms that the largest deviation in the measured observables occurs in $P_5^\prime$.
 
\begin{table}[!b]
\centering
\begin{tabular}{|c|c|c|c|c|c|c|}\hline
& $d{\cal B}/dq^2$ & $F_L$ & ${\cal A}_{FB}$ & $P_4^{\prime}$ & $P_5^{\prime}$ & $S_3$ \\ \hline
$\chi^2$ & $0.49$ & $0.19$ & $0.79$  & $0.91$ & $2.09$& $0.53$\\
\hline
\end{tabular}
\caption{$\chi^2$ values for the various observables, defined as  $\dd \chi^2= \frac{1}{N} \sum_i^N\frac{ \left({\cal O}^{\text{exp}}_i-{\cal O}^{\text{th}}_i\right)^2}{\delta_i^2+\sigma_i^2}$.  $i$ runs over the $N$ experimental bins, ${{\cal O}^{\text{exp}}_i}$ and ${\cal O}^{\text{th}}_i$ are the LHCb measurements and the theoretical results in the RS$_c$ model, with errors $\delta_i$ and $\sigma_i$, respectively.}
\label{tab:chi2}
\end{table}

\begin{figure}[h]
\includegraphics[width = 0.4\textwidth]{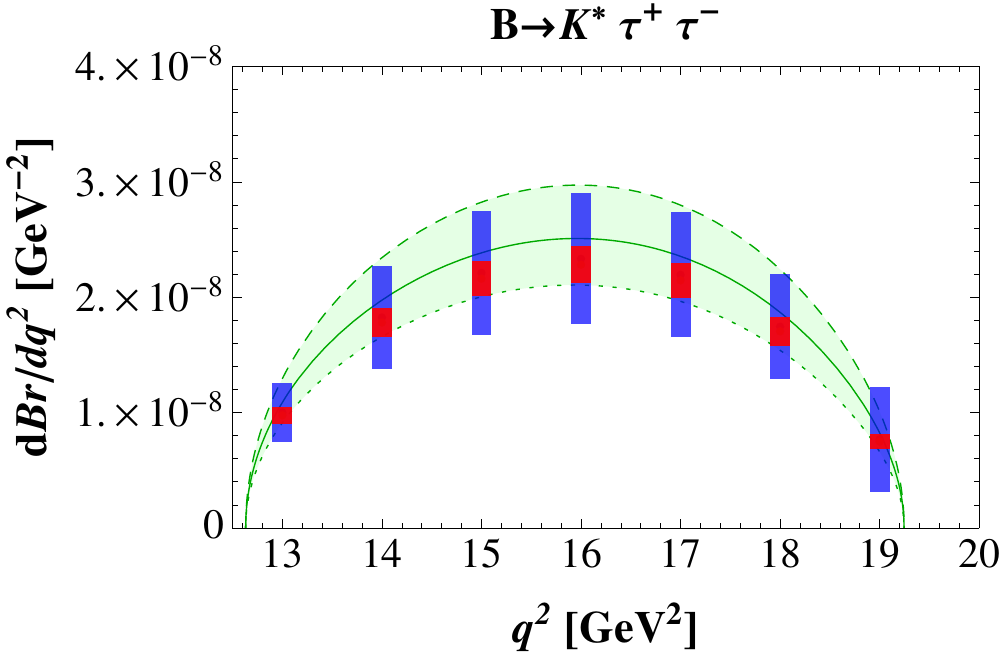}
\caption{Differential $B\to K^* \tau^+ \tau^-$ decay rate. The light (green) band corresponds to the SM result, including the uncertainty in the form factors. The red and blue vertical bars correspond to the RS$_c$ result,  without or with  the uncertainty in form factors.}\label{fig:spectrtau}
\end{figure}

\begin{figure}[h]
\includegraphics[width = 0.4\textwidth]{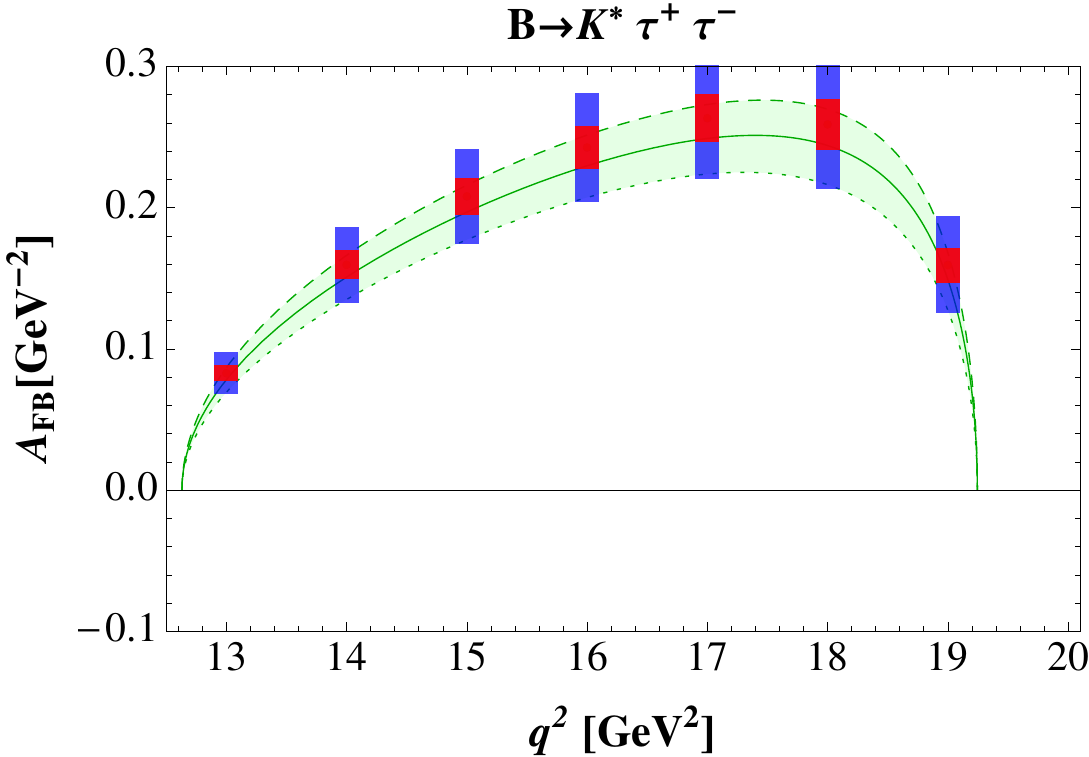}
\caption{ Lepton FB asymmetry in  $B\to K^* \tau^+ \tau^-$. The symbols have the same meaning  as in Fig.~\ref{fig:spectrtau}.}\label{fig:AFBtau}
\end{figure}

\begin{figure}[h]
\includegraphics[width = 0.4\textwidth]{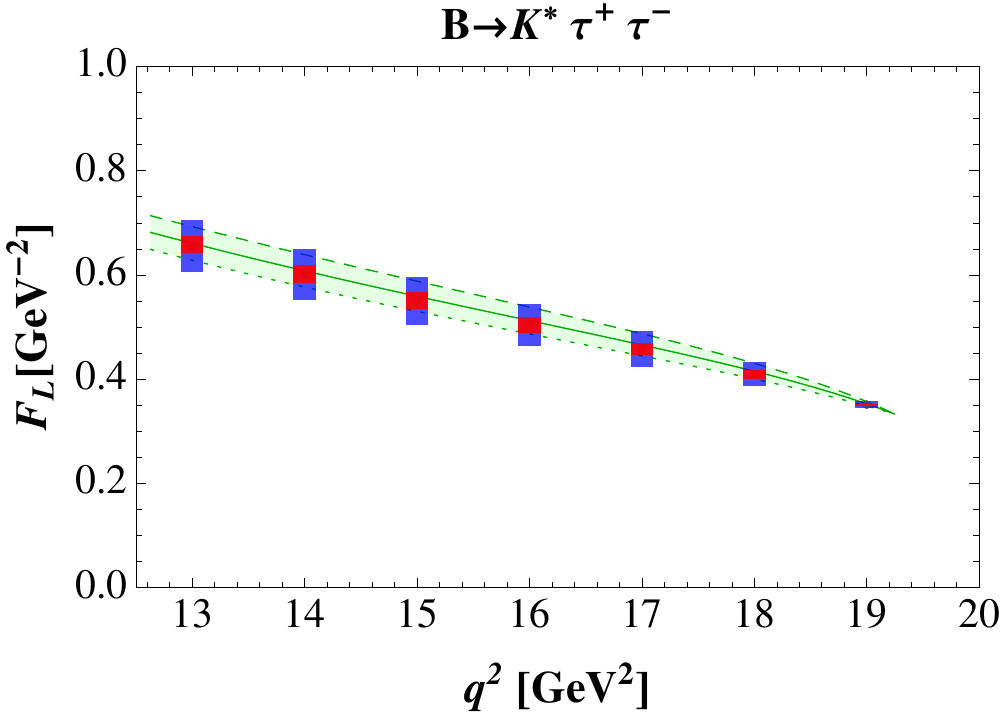}
\caption{$K^*$ longitudinal polarization fraction in  $B\to K^* \tau^+ \tau^-$. The symbols have the same meaning  as in Fig.~\ref{fig:spectrtau}. }\label{fig:FLtau}
\end{figure}

\begin{figure}[h]
\includegraphics[width = 0.4\textwidth]{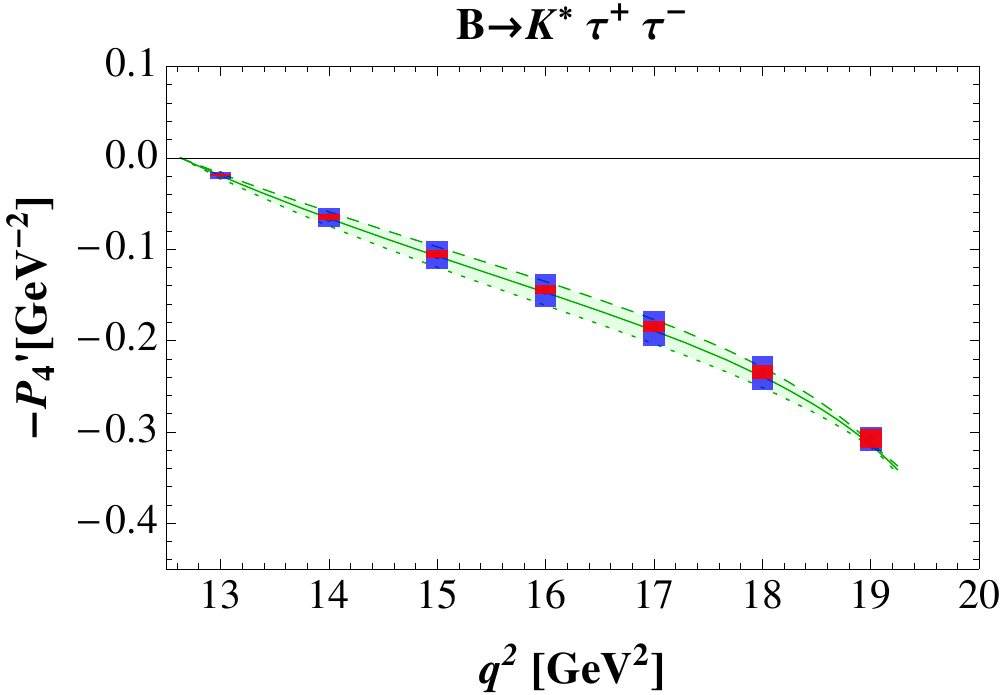}
\caption{Observable $P^\prime_4$ in  $B\to K^* \tau^+ \tau^-$. The symbols have the same meaning  as in Fig.~\ref{fig:spectrtau}. The sign is fixed to make  compatible the definition (\ref{Piprime}) and the one adopted in Ref. \cite{Aaij:2013qta}, as in the case of $\mu^+ \mu^-$. }\label{fig:P4ptau}
\end{figure}

\begin{figure}[h]
\includegraphics[width = 0.4\textwidth]{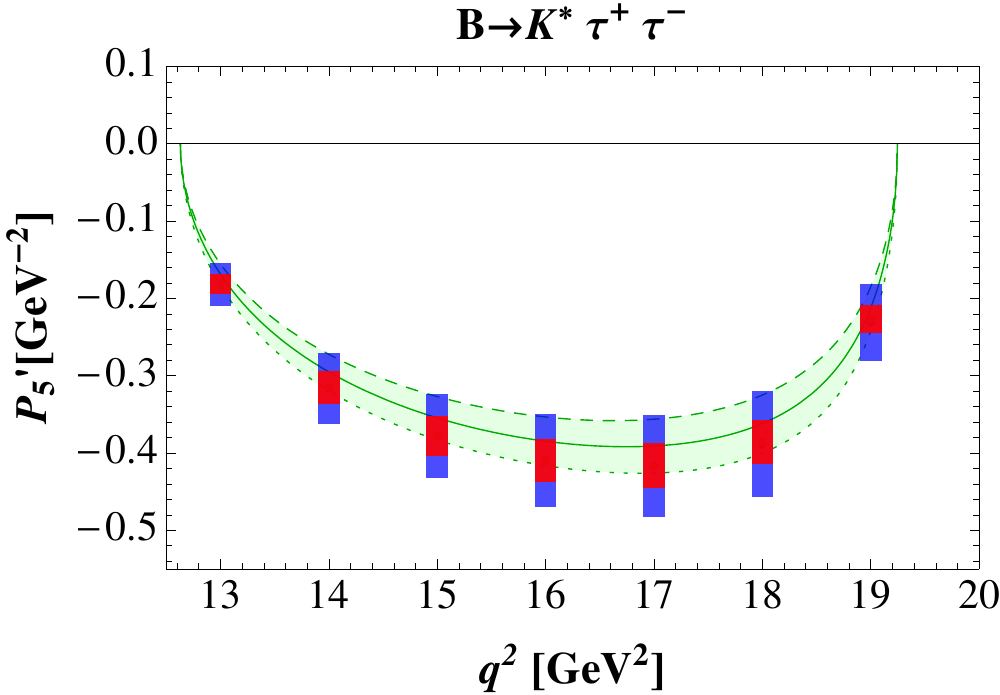}
\caption{Observable $P^\prime_5$ in  $B\to K^* \tau^+ \tau^-$. The symbols have the same meaning  as in Fig.~\ref{fig:spectrtau}.}\label{fig:P5ptau}
\end{figure}

\begin{figure}[h]
\includegraphics[width = 0.4\textwidth]{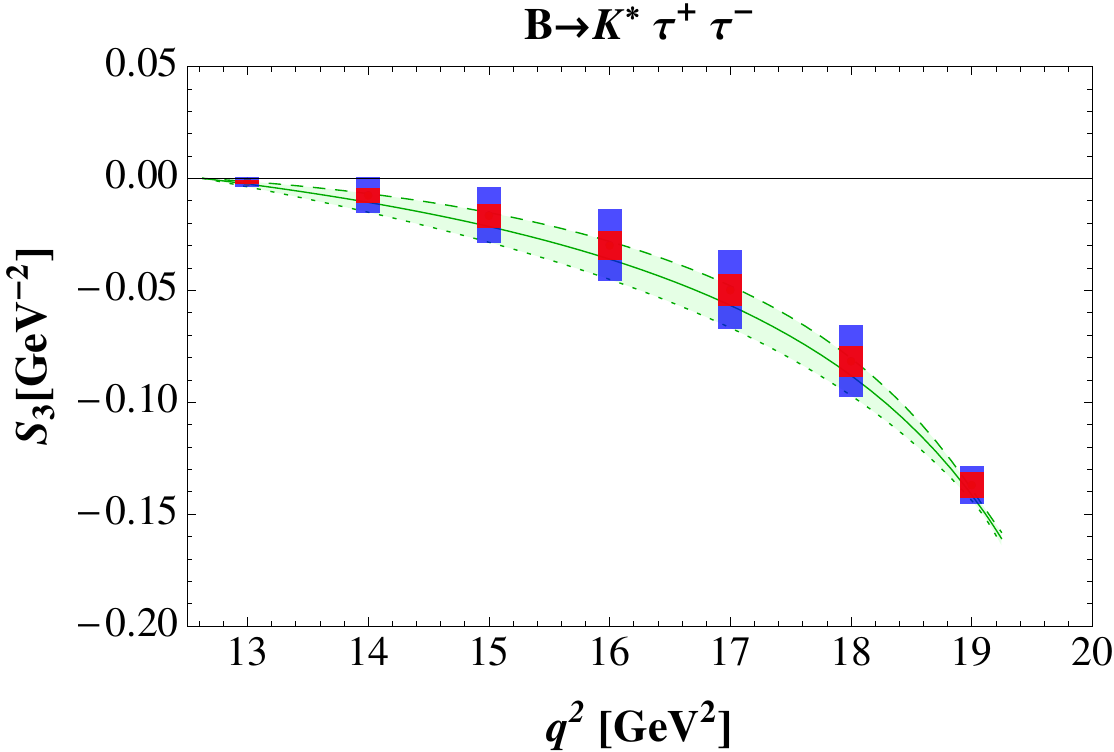}\\
\caption{Observable $S_3$  in  $B\to K^* \tau^+ \tau^-$. The symbols have the same meaning  as in Fig.~\ref{fig:spectrtau}.}\label{fig:S3tau}
\end{figure}

\begin{figure}[h]
\includegraphics[width = 0.4\textwidth]{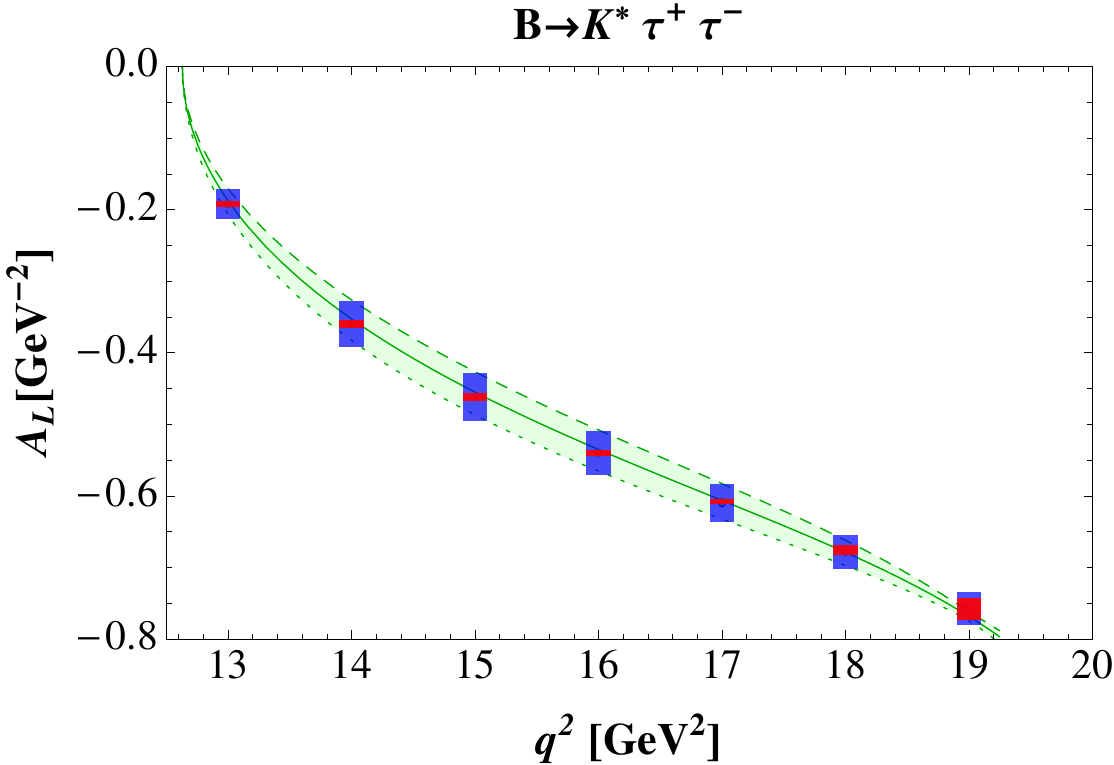}\\ \vspace*{0.5cm}
\includegraphics[width = 0.4\textwidth]{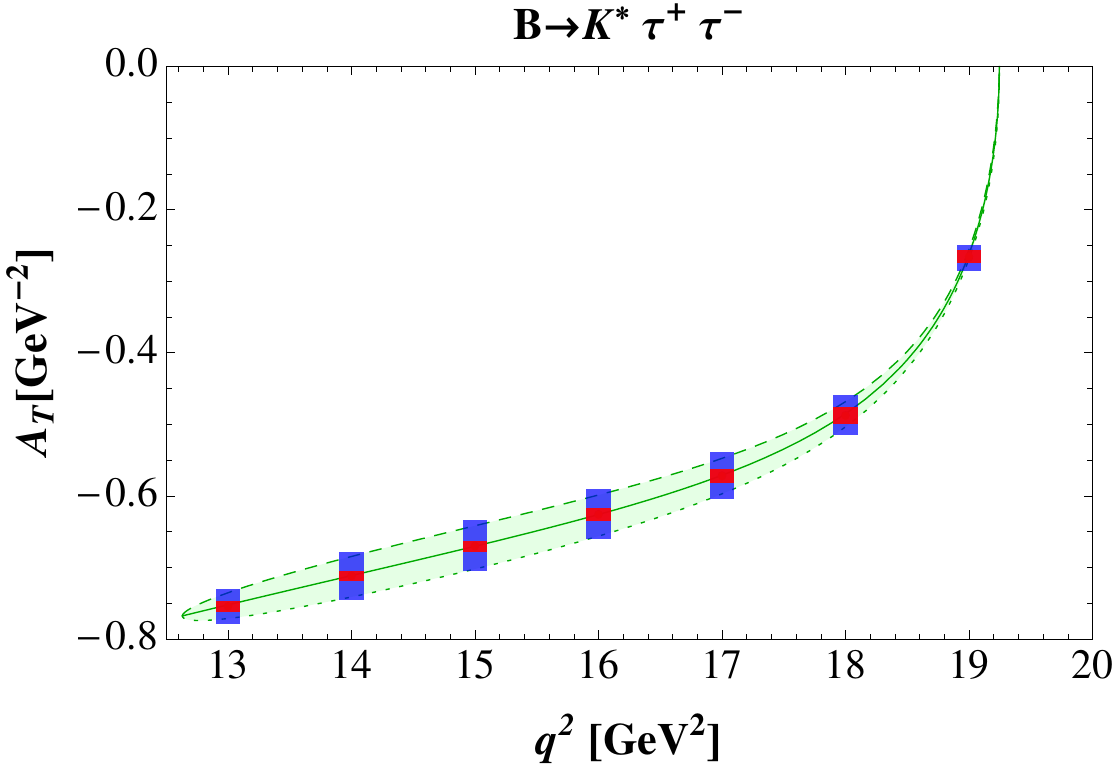}
\caption{Longitudinal  and transverse $\tau$ polarization asymmetries. The symbols have the same meaning  as in Fig.~\ref{fig:spectrtau}.}\label{fig:ALtau}
\end{figure}

The results for the case of $\tau^+ \tau^-$  final state are shown in Figs.~\ref{fig:spectrtau}-\ref{fig:S3tau}. The kinematically accessible $q^2$ range starts at $q^2 \simeq 12.628$ GeV$^2$,
so that the small effects  in the muon mode  at low $q^2$ do not appear in this case. In the decay rate distribution and in the lepton FB asymmetry the
results in RS$_c$ systematically deviate from SM in the full parameter space, but the effect is smaller than the hadronic uncertainty. Such a systematic deviation  also appears in 
$P^\prime_5$ and  $S_3$, Fig.\ref{fig:P5ptau} and \ref{fig:S3tau}, while the longitudinal and transverse $\tau$ polarization asymmetries essentially coincide with the SM ones,
Fig.\ref{fig:ALtau},
%and \ref{fig:ATtau}) 
and seem not suitable for characterizing the considered new physics model. 

\begin{figure}[h]
\includegraphics[width = 0.4\textwidth]{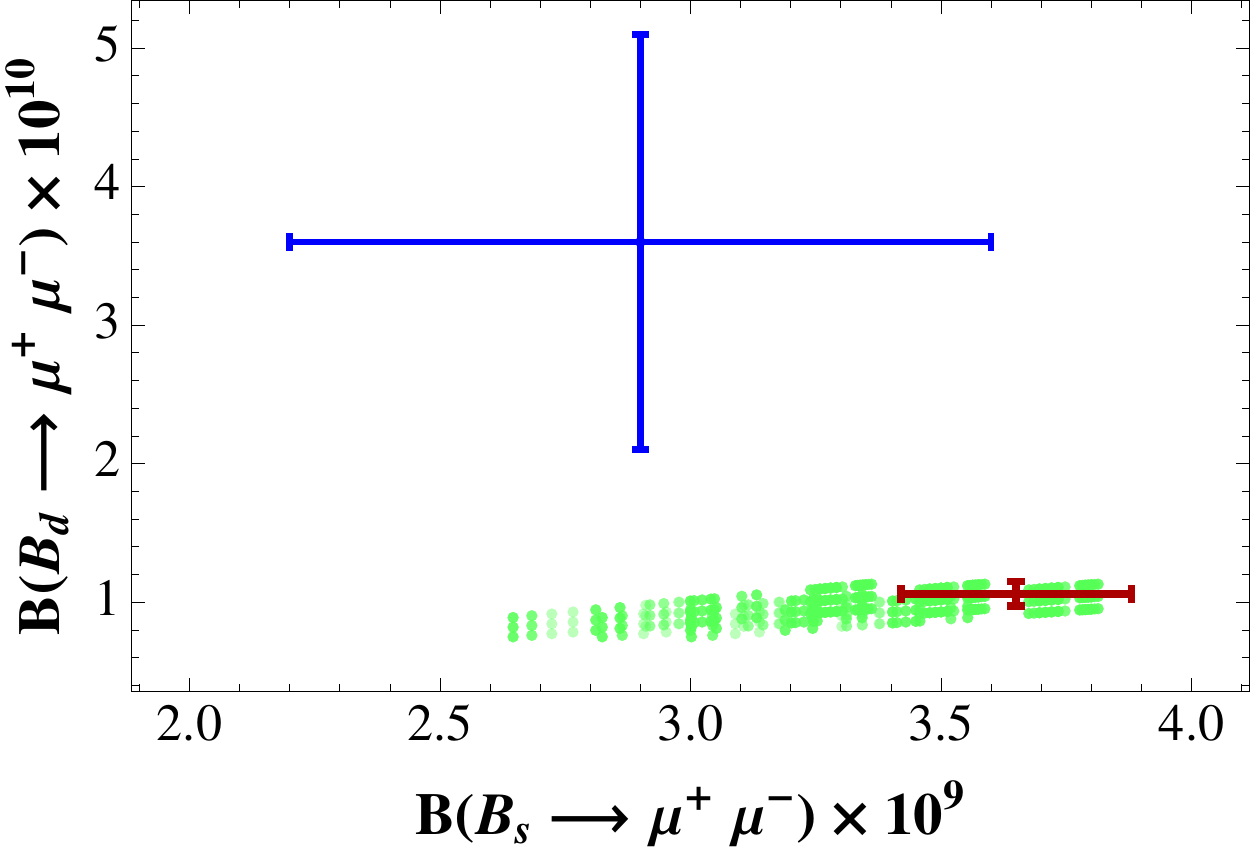}
\caption{Correlation between the branching ratios of $B_s \to \mu^+ \mu^-$ and $B_d\to \mu^+ \mu^-$. The blue bars represent the experimental data  (\ref{bsdmumuexp}),  the red ones  the SM predictions (\ref{bsdmumuSM}).
The green region is the prediction in RS$_c$ derived in this paper.}\label{fig:bsdmumu}
\end{figure}

A final remark concerns the rare $B_{s,d}\to \mu^+ \mu^-$ modes. 
As  stated in the Introduction, an important breakthrough  is the measurement in Eq.~(\ref{bsdmumuexp}).
Since the theoretical predictions for the branching fractions of these modes depend on a subset of the Wilson coefficients that have been considered, we can derive predictions  using the same  set of parameters, see Fig.~\ref{fig:bsdmumu}.
The SM result depends only on the coefficient $C_{10}$, and in   RS one has to replace $C_{10} \to C_{10}-C_{10}^\prime$.
In Fig.~\ref{fig:bsdmumu} we show the correlation between the two modes, comparing the RS prediction to  SM and to the experimental data.
There is a region of the parameter space in which the SM result for both branching ratios is reproduced. However, the allowed range  in RS$_c$ is larger than in SM: ${\cal B}(B_s \to \mu^+ \mu^-)|_{RS} \in [2.64,\,3.83]\,\times 10^{-9}$ and ${\cal B}(B_d \to \mu^+ \mu^-)|_{RS} \in [0.70,\,1.16]\,\times 10^{-10}$, in the right direction in the case of $B_s$ when comparing with data, but still lower than the  datum for  $B_d$. A similar result was already found in \cite{buras2}, although with a smaller deviation with respect to SM, in particular in the $B_s$ case.
%This correlation had already been analyzed in the framework of RS$_c$ in \cite{buras2}:  the result of that analysis  is that the branching fractions of $B_s \to \mu^+ \mu^-$ and $B_d\to \mu^+ \mu^-$ may assume values both above and below the SM central values, a difference which likely lies in the set of chosen parameters. 

\section{Conclusions}\label{sec:conc}
We have studied several observables of the rare decays $B \to K^* \ell^+ \ell^-$ within the RS$_c$ model, in the case where measurements are available and in the case of massive leptons.   We have carried out a new calculation of the
contributions to the Wilson coefficients $C^{(\prime)}_{7,8}$ and a new scan of the model parameters, imposing the experimental  constraints of CKM and quark mass values.
The obtained deviations with respect SM are small, and at present they are generally hidden by the uncertainties  on the hadronic form factors for several observables. However, in few cases the deviations from SM are systematic in the full $q^2$ range, and are similar to the 
presently accepted hadronic uncertainties: this renders  such observables of great interest in view of searching  signals of this possible extension of the Standard Model.

\acknowledgements
We thank A.J. Buras for fruitful discussions. We also thank E. Scrimieri for suggestions useful for the numerical analysis.

\appendix\label{app}
\section{Gauge boson and fermion profiles} \label{app:profiles}
The gauge boson and fermion profiles, entering  in the calculations carried out   in this paper,  are  obtained solving the equations of motion for the corresponding fields \cite{Davoudiasl:1999tf,gherghetta}:
\begin{itemize}
\item 
{\bf profile of the first KK excitation of a gauge boson having a zero mode}
\be
g(y)=\frac{e^{ky}}{N_1}\left[J_1 \left( \frac{m_1}{k} e^{ky}\right)+b_1(m_1) Y_1 \left( \frac{m_1}{k} e^{ky}\right)\right] \,\,\, , 
\ee
with  $J_1$ and $Y_1$   Bessel functions,
\bea
b_1(m_1)&=&-\frac{J_1 \left( \frac{m_1}{k}\right) +\frac{m_1}{k} J_1^\prime \left( \frac{m_1}{k}\right)}{Y_1 \left( \frac{m_1}{k}\right) +\frac{m_1}{k} Y_1^\prime \left( \frac{m_1}{k}\right)} \,\,\, ,  \\
N_1 &=& \frac{e^{kL/2}}{\sqrt{\pi L m_1}} \,\,\, . 
\eea
We use $m_1 \simeq 2.45\, f$.
\item 
{\bf profile of the first KK level of a gauge boson without  zero mode}
\be
{\tilde g}(y)=\frac{e^{ky}}{N_1}\left[J_1 \left( \frac{\tilde m_1}{k}e^{ky}\right)+\tilde b_1(\tilde m_1) Y_1 \left( \frac{\tilde m_1}{k}e^{ky}\right)\right]
\ee
where now 
\be
\tilde b_1(\tilde m_1)=-\frac{J_1 \left( \frac{\tilde m_1}{k}\right) }{Y_1 \left( \frac{\tilde m_1}{k}\right) }  \,\,\, .
\ee
We use $\tilde m_1 \simeq 2.40\, f$.
\item 
{\bf profile of the fermion zero mode}
\be
f^{(0)}(y,c)=\sqrt{\frac{(1-2c)kL}{e^{(1-2c)kL}-1}}e^{-cky} \,\,.
\ee
\end{itemize}

\section{Feynman rules for neutral current interactions} \label{app:feynman}

\begin{figure}[b!]
\centering
\includegraphics[width = 0.35\textwidth]{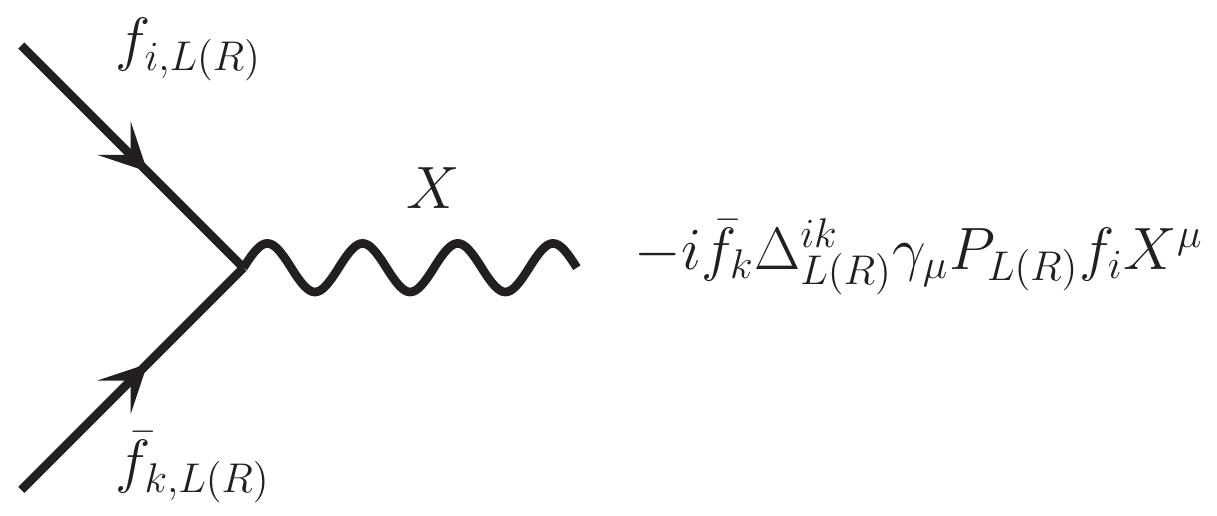}
\caption{Couplings of neutral gauge bosons $X=Z, \, Z_H,\,Z^\prime,\,A^{(1)}$ to fermions $f$ with flavour indices $i,k$.
In addition to flavour conserving couplings, $i=k$, flavour violating couplings $i \neq k$ are possible.}\label{fig:FR}
\end{figure}

We need to consider the neutral current interactions mediated by the gauge bosons $Z, \, Z_H,\,Z^\prime$ and $A^{(1)}$.  As  mentioned,  such interactions can be either flavour conserving or flavour violating.  We collect  in four triplets the up-type quarks, the down-type quarks, the neutrinos and the charged leptons: $f=\left( \begin{array}{c} f_1 \\ f_2 \\ f_3 \end{array}\right),$ with $f=u_i,\,d_i,\,\nu_{\ell_i},\ell_i^-$,  and  $i=1,2,3$ a generation index.
We define the effective coupling of a generic gauge boson $X$ to a pair of fermions $f_i\,f_k$:
\be
{\cal L}_{int}^X=-i X^\mu \left( \Delta_{L}^{f_i f_k}{\bar f}_{k,L} \gamma_\mu P_L f_{i,L}+\Delta_{R}^{f_i f_k}{\bar f}_{k,L} \gamma_\mu P_R f_{i,L} \right) \,\,, \label{neutral}
\ee
where $\dd P_{L,R}=\frac{1 \mp \gamma_5}{2}$. Fig.~\ref{fig:FR}  shows this vertex.
In  the expression of the couplings $\Delta_{L,R}^{f_i f_k}$,   two more overlap integrals are needed:
\bea
I_1^+ &=& \frac{1}{L} \int_0^L dy \,   e^{-2ky} [h(y)]^2 \, g(y) \,\, , \nn \\
I_1^- &=& \frac{1}{L} \int_0^L dy  \,   e^{-2ky} [h(y)]^2 \, {\tilde g}(y)  \,\, , 
 \eea
together with  the quantity
\be
D^{f_i f_k}=\tilde{\cal R}_{f_i f_k}I_1^- - {\cal R}_{f_i f_k} I_1^+ \,\,.
\ee
$D^f$ is the diagonal matrix with elements  $D^{f_i f_i}$.
We consider  separately the case of the four neutral gauge bosons listed above. In the following, 
 ${\cal M}$ indicates  one of the matrices  ${\cal U}_L$, ${\cal U}_R$, ${\cal D}_L$, ${\cal D}_R $, for up or down-type left-  and right-handed quarks. The obtained rules also hold  in the case of leptons with ${\cal M}$ being the unit matrix.
In the case of the couplings of $Z$, $Z^\prime$ and $Z_H$ the Feynman rules are obtained expanding in the small parameter 
$\dd \epsilon=\frac{g^2 v^2}{4L M^2}$, with $g$ the $5D$ $SU(2)_L$ gauge constant and  $M^2=(m_1^2+{\tilde m}_1^2)/2$.
Since, neglecting corrections of ${\cal O}(v^2/M^2)$ the mixing angle $\psi$ in (\ref{1stmix}-\ref{cpsi}) coincides with the Weinberg angle $\theta_W$, in the following Feynman rules we put $s_\psi=s_W=\sin \theta_W$ and $c_\psi=c_W=\cos \theta_W$.

\begin{itemize}
\item  {\bf Couplings of the photon 1-mode $A^{(1)}$}
\par \noindent
The couplings are given by
\be
\Delta_L^{f_i f_k}(A^{(1)})=\Delta_R^{f_i f_k}(A^{(1)})=Q_f \, e \left({\cal M}^\dagger {\cal R}_f {\cal M} \right)_{ki}
\ee
where $Q_f$ is the electric charge of the fermion $f$ (in units of the positron charge $e$). \\

\item  {\bf $Z$ couplings }
\par \noindent
In terms of  the SM couplings of the $Z$ boson:
\be
[g_Z^{SM}(f)]=-\frac{e}{s_W c_W}\left(T_3 -s_W^2 Q_f \right)  \,\,\, ,
\ee
with  $T_3$   the third component of the weak $SU(2)_L$  isospin,
we find:
\bea
\Delta^{f_i f_k}(Z) \hspace*{5cm}\nn \\=[g_Z^{SM}(f)] \left\{ \delta_{ki}+ \frac{\epsilon}{c^2_W} \left({\cal M}^\dagger D^f{\cal M} \right)_{ki} \right\}
\eea
for $f=u_R,\,c_R,\,t_R$, for  $f=d_L,\,s_L,\,b_L$,  and for $f=\ell^-_L$,
and
\bea
\Delta^{f_i f_k}(Z) \hspace*{5cm}\nn \\
=[g_Z^{SM}(f)] \left\{ \delta_{ki}  + \frac{\epsilon}{c^2_W}\left({\cal M}^\dagger D^f{\cal M} \right)_{ki}\right\}
\nn \\
+ \frac{\epsilon e}{s_Wc_W} I_1^- \left({\cal M}^\dagger \tilde{\cal R}_f{\cal M}\right)_{ki} \hspace*{2.2cm}
\eea
for
$f=u_L,c_L,t_L$, for  $f=d_R,s_R,b_R$,  and for $f=\ell^-_R,\nu_{e,\,\mu,\,\tau}$.
\item {\bf $Z_H$ couplings} 
\par
At the order ${\cal O}(\epsilon^0)$ for the heavier bosons $Z_H$ and $Z^\prime$ we find:
\be
\Delta^{f_i f_k}(Z_H) 
= c_\xi (\Delta_f^c)_{ki}(Z_H)+s_\xi (\Delta_f^s)_{ki}(Z_H)\, ,\label{ZH-coup}
\ee
where $c_\xi=\cos (\xi)$ and $s_\xi=\sin(\xi)$, and
\bea
(\Delta_f^c)_{ij}(Z_H)&=&g_Z^{SM}(f) \left( {\cal M}^\dagger {\cal R}_f {\cal M} \right)_{ij} \,\,\, , \nn \\
(\Delta_f^s)_{ij}(Z_H)&=&\frac{\left( {\cal M}^\dagger \tilde{\cal R}_f {\cal M} \right)_{ij} }{\sqrt{1-2s_W^2}} \left[ g_Z^{SM}(f) +g c_W \right] , 
\eea
for $f=u_L,c_L,t_L$,  and for $f=d_R,s_R,b_R$,  and
\bea
(\Delta_f^c)_{ij}(Z_H)&=&  g_Z^{SM}(f) \left( {\cal M}^\dagger {\cal R}_f {\cal M} \right)_{ij} \,\,\, , \nn \\
(\Delta_f^s)_{ij}(Z_H)&=&\frac{\left( {\cal M}^\dagger \tilde{\cal R}_f {\cal M} \right)_{ij} }{\sqrt{1-2s_W^2}}  g_Z^{SM}(f) \,\,\, , 
\eea
for $f=u_R,c_R,t_R$,  and for $f=d_L,s_L,b_L$.
For leptons we have:
\bea
\Delta^{\ell \ell}(Z_H) \hspace*{5.5cm} \nn \\ 
=g_Z^{SM}(\ell) \left[c_\xi \Delta^{c,\ell \ell}(Z_H)+s_\xi\Delta^{s,\ell \ell}(Z_H) \right]\,, \label{ZH-coup-lept}
\eea
where
\bea
\Delta^{c,\ell \ell}(Z_H)&=& {\cal R}_{\ell \ell} \hskip 1.5 cm (\ell=\nu,\ell^-_L,\ell^-_R) \nn \\
\Delta^{s,\nu \nu}(Z_H)&=&-\tilde{\cal R}_{\nu \nu} \sqrt{1-2 s_W^2} \nn \\
\Delta^{s,\ell^-_L \ell^+_L}(Z_H)&=&\frac{\tilde{\cal R}_{\ell^-_L \ell^+_L}}{ \sqrt{1-2 s_W^2}}  \\
\Delta^{s,\ell^-_R \ell^+_R}(Z_H)&=&-\tilde{\cal R}_{\ell^-_R \ell^+_R}\frac{ \sqrt{1-2 s_W^2}}{s^2_W} \,\,\, . \nn
\eea
\item  {\bf $Z^\prime$ couplings }
\par
Defining the couplings with the same structure as in Eq.(\ref{ZH-coup}), we find:
\bea
(\Delta_f^c)_{ij}(Z^\prime)&=&(\Delta_f^s)_{ij}(Z_H)
\nn \\
(\Delta_f^s)_{ij}(Z^\prime)&=&-(\Delta_f^c)_{ij}(Z_H) \,\,\, .
\eea
For  leptons we have
\bea
\Delta^{c,\ell \ell}(Z^\prime)&=&\Delta^{s,\ell \ell}(Z_H)
\nn \\
\Delta^{s,\ell \ell}&=&-\Delta^{c,\ell \ell}(Z_H) \,\,\, . 
\eea
\end{itemize}

\section{Calculation of $\Delta C_{7,8}^{(\prime)}$ at the high scale $M_{KK}$}\label{app:c7}

\begin{figure*}[ht!]
\begin{tabular}{ccc}
\includegraphics[width = 0.2\textwidth]{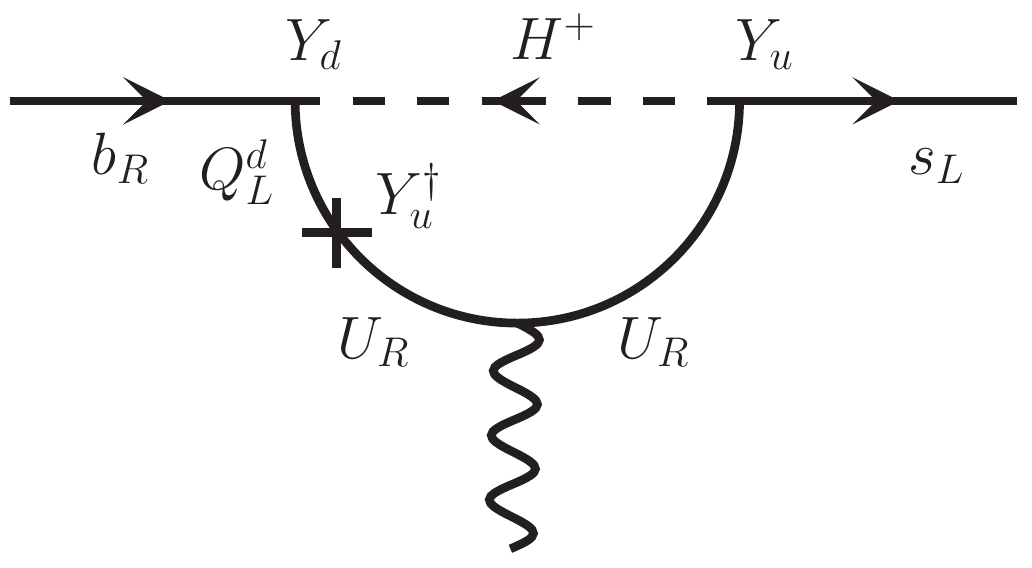} & \includegraphics[width = 0.2\textwidth]{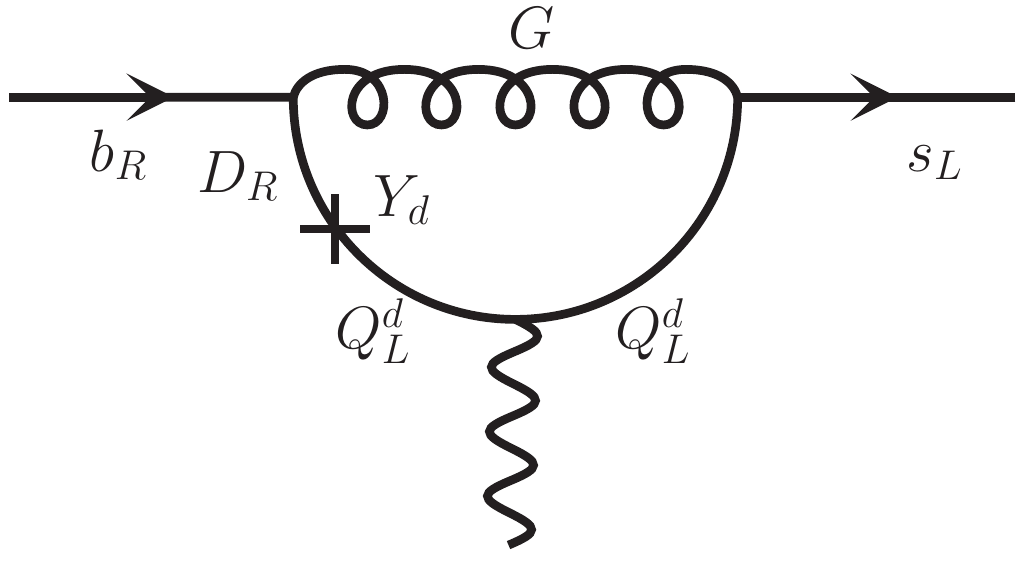}& \includegraphics[width = 0.2\textwidth]{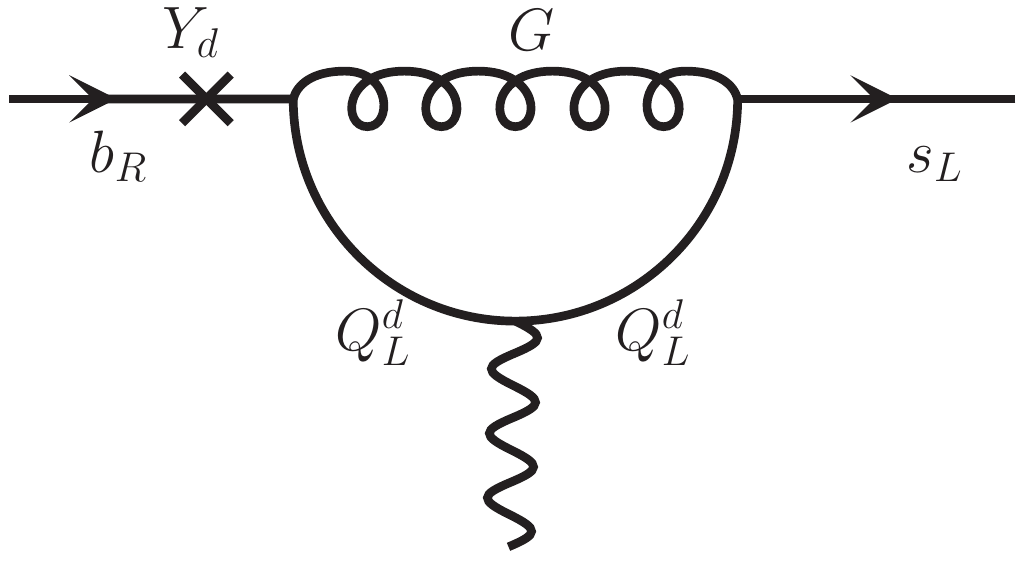}\\$(\Delta C_7^{(\prime)})_1$ & $(\Delta C_7^{(\prime)})_2$ & $(\Delta C_7^{(\prime)})_3$
\\
\includegraphics[width = 0.2\textwidth]{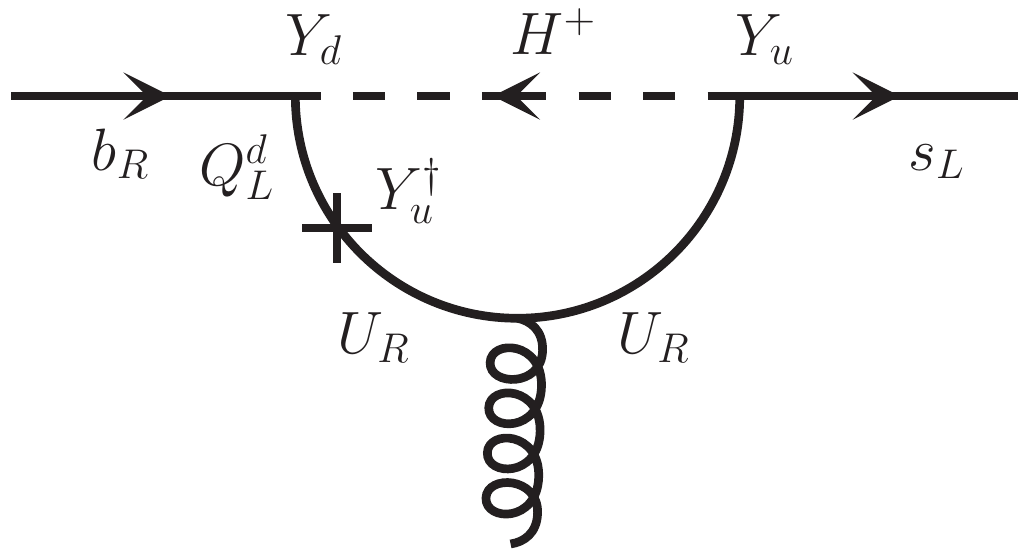} & \includegraphics[width = 0.2\textwidth]{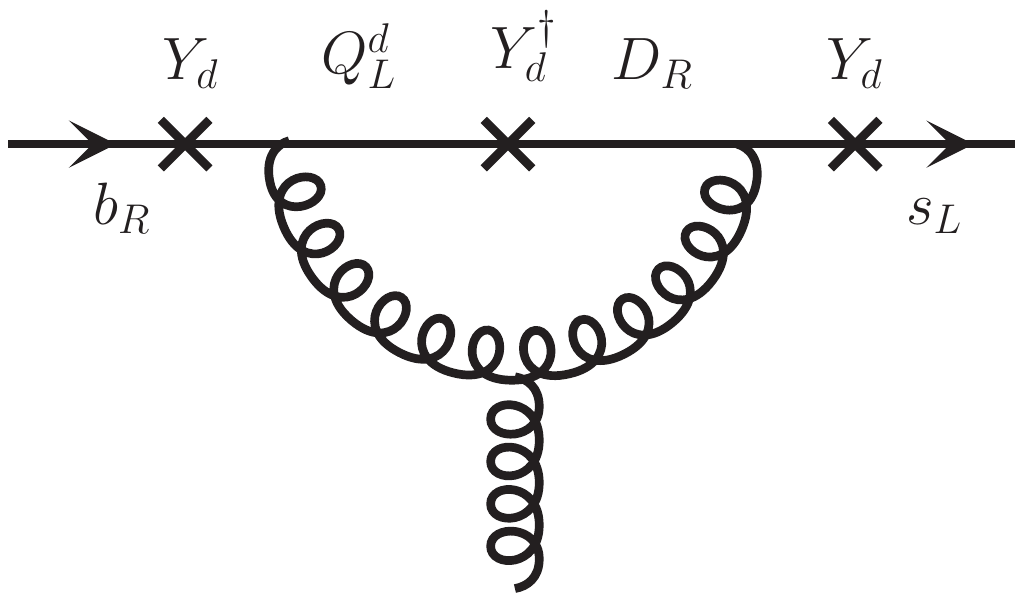}& \includegraphics[width = 0.2\textwidth]{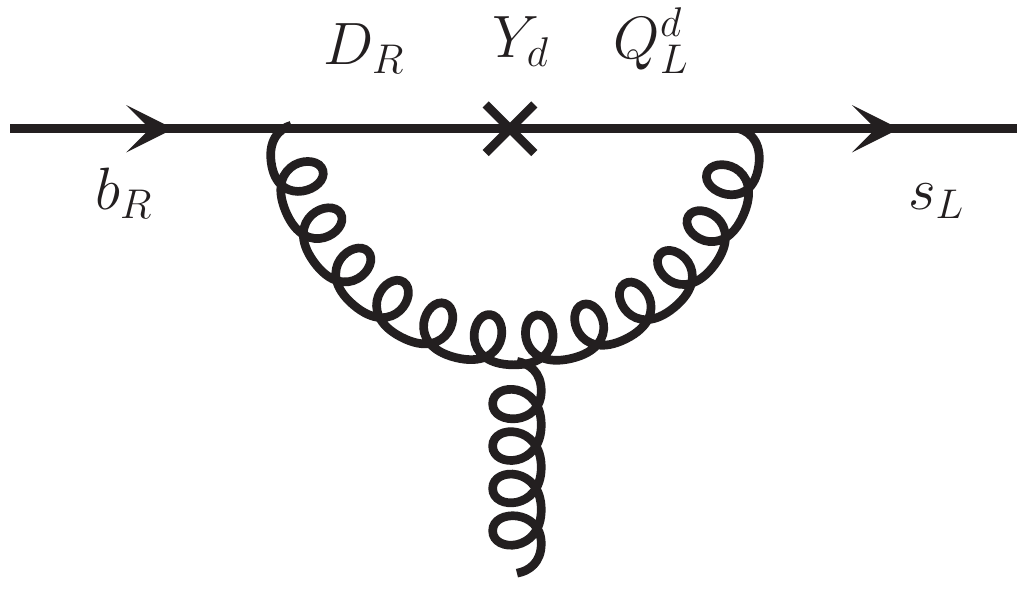}\\$(\Delta C_8^{(\prime)})_1$ & $(\Delta C_8^{(\prime)})_2$ & $(\Delta C_8^{(\prime)})_3$
\\
\end{tabular}
\caption{New penguin diagrams contributing to the Wilson coefficients $C_{7,8}^{(\prime)}$ in the RS$_c$ model. Diagrams with specular mass insertions must also be considered.}\label{fig:diagC7}
\end{figure*}

To compute $\Delta C_{7,8}^{(\prime)}$ at the scale $M_{KK}$ one has to consider penguin diagrams with new particles in addition to the SM ones. In the approximation adopted in this paper of neglecting higher KK modes for fermions, the dominant diagrams (shown in Fig.~\ref{fig:diagC7})  are of the following type \cite{Csaki:2010aj,Blanke:2012tv}:
\begin{itemize}
\item
Penguin diagrams mediated by a charged Higgs in which a photon (for $C_7^{(\prime)}$) or a gluon (for $C_8^{(\prime)}$) is emitted from an internal up-type quark, with a mass insertion on the internal fermion line.
% is necessary to reproduce the right chirality for external quarks. 
%Therefore, two such diagrams are present in which one symmetrizes the position of the mass insertion with respect to the photon vertex. 
Diagrams contributing to the primed coefficients differ from those for the unprimed  for the chirality of the external quarks. 
%The contributions   $(\Delta C_7^{(\prime)})_1$  and  $(\Delta C_8^{(\prime)})_1$ in Fig.~\ref{fig:diagC7} are obtained.
\item
Penguin diagrams mediated by  a gluon with an internal  down-type quark. 
%These are different for $C_7^{(\prime)}$ and $C_8^{(\prime)}$. In the case of
For $C_7^{(\prime)}$ we have the  contributions $(\Delta C_7^{(\prime)})_2$ and  $(\Delta C_7^{(\prime)})_3$  as in Fig.~\ref{fig:diagC7},
% where again the  two symmetric possibilities for the mass insertion should be considered.
 for $C_8^{(\prime)}$  the  contributions $(\Delta C_8^{(\prime)})_2$ and  $(\Delta C_8^{(\prime)})_3$ involve the three-gluon vertex. 
\end{itemize}
We list below the results. 

\begin{widetext}
%\widetext
\bea
(\Delta C_7)_1 &=& i\, Q_u\,r \, \sum_{F=u,c,t}\left[A+2m_F^2(A^\prime+B^\prime) \right] \left[{\cal D}_L^\dagger Y^u (Y^u)^\dagger Y^d {\cal D}_R \right]_{23} \nn \\
(\Delta C_7)_2 &=&- i\, Q_d\,r \frac{8}{3} (g_s^{4D})^2 \, \sum_{F=d,s,b}\left[I_0+A+B+4m_F^2(I_0^\prime+A^\prime+B^\prime) \right] \left[{\cal D}_L^\dagger {\cal R}_L Y^d {\cal R}_R{\cal D}_R \right]_{23}  \\
(\Delta C_7)_3 &=&i\, Q_d\,r \frac{8}{3} (g_s^{4D})^2 \, \sum_{F=d,s,b} \frac{m_F}{m_b} \,  \left[I_0+A+B \right] \left\{ \left[{\cal D}_L^\dagger {\cal R}_L {\cal R}_L Y^d {\cal D}_R \right]_{23} +  \frac{m_b}{m_s} \left[{\cal D}_L^\dagger  Y^d {\cal R}_R {\cal R}_R {\cal D}_R \right]_{23} \right\} \nn \eea
\bea
(\Delta C_7^\prime)_1 &=& i\, Q_u\,r \, \sum_{F=u,c,t}\left[A+2m_F^2(A^\prime+B^\prime) \right] \left[{\cal D}_R^\dagger  (Y^d)^\dagger Y^u (Y^u)^\dagger {\cal D}_L \right]_{23} \nn \\
(\Delta C_7^\prime)_2 &=&- i\, Q_d\,r \frac{8}{3} (g_s^{4D})^2 \, \sum_{F=d,s,b}\left[I_0+A+B+4m_F^2(I_0^\prime+A^\prime+B^\prime) \right] \left[{\cal D}_R^\dagger {\cal R}_R (Y^d)^\dagger {\cal R}_L {\cal D}_L \right]_{23}  \\
(\Delta C_7^\prime)_3 &=&i\, Q_d\,r \frac{8}{3} (g_s^{4D})^2 \, \sum_{F=d,s,b} \frac{m_F}{m_b} \, \left[I_0+A+B \right] \left\{ \left[{\cal D}_R^\dagger {\cal R}_R {\cal R}_R (Y^d)^\dagger {\cal D}_L \right]_{23} +  \frac{m_b}{m_s} \left[{\cal D}_R^\dagger  (Y^d)^\dagger {\cal R}_L {\cal R}_L {\cal D}_L \right]_{23} \right\} \nn \eea
\bea
(\Delta C_8)_1 &=& i\, r \, \sum_{F=u,c,t}\left[{A}+2m_F^2({ A}^\prime+{B}^\prime) \right] \left[{\cal D}_L^\dagger Y^u (Y^u)^\dagger 
Y^d {\cal D}_R \right]_{23} \nn \\
(\Delta C_8)_2 &=&- i\, r \frac{9}{8} (g_s^{4D})^2 \, \frac{v^2}{m_b \, m_s}  \,{\cal T}_3 \, \sum_{F=d,s,b}\left[{\bar A}+{\bar B}+2m_F^2({\bar A}^\prime+{\bar B}^\prime) \right] \left[{\cal D}_L^\dagger Y^d {\cal R}_R (Y^d)^\dagger {\cal R}_L Y^d {\cal D}_R \right]_{23}  \\
(\Delta C_8)_3 &=&-i\, \,r \,\frac{9}{4} (g_s^{4D})^2 \, {\cal T}_3 \,\sum_{F=d,s,b}\left[{\bar A}+{\bar B} +2m_F^2({\bar A}^\prime+{\bar B}^\prime) \right]  \left[{\cal D}_L^\dagger {\cal R}_L Y^d  {\cal R}_R {\cal D}_R \right]_{23} \nn \eea
\bea
(\Delta C_8^\prime)_1 &=& i\, r \, \sum_{F=u,c,t}\left[{ A}+2m_F^2({A}^\prime+{ B}^\prime) \right] \left[{\cal D}_R^\dagger  (Y^d)^\dagger Y^u (Y^u)^\dagger {\cal D}_L \right]_{23} \nn \\
(\Delta C_8^\prime)_2 &=&- i\, r \frac{9}{8} (g_s^{4D})^2 \, \frac{v^2}{m_b\, m_s }  \,{\cal T}_3 \, \sum_{F=d,s,b}\left[{\bar A}+{\bar B}+2m_F^2({\bar A}^\prime+{\bar B}^\prime) \right] \left[{\cal D}_R^\dagger (Y^d)^\dagger {\cal R}_L Y^d {\cal R}_R (Y^d)^\dagger {\cal D}_L \right]_{23}  \\
(\Delta C_8^\prime)_3 &=&-i\, \,r \,\frac{9}{4} (g_s^{4D})^2 \, {\cal T}_3 \,\sum_{F=d,s,b}\left[{\bar A}+{\bar B} +2m_F^2({\bar A}^\prime+{\bar B}^\prime) \right]  \left[{\cal D}_R^\dagger {\cal R}_R (Y^d)^\dagger  {\cal R}_L {\cal D}_L \right]_{23} \,\,\, . \nn \eea
\end{widetext}
We have defined $\dd r=\displaystyle{\frac{v}{\frac{G_F}{4 \pi^2}\, V_{tb} \, V_{ts}^* \, m_b}}$ and ${\cal T}_3$ is the overlap of the  profiles of two KK 1-mode and one KK 0-mode gluons:
$\dd {\cal T}_3=\frac{1}{L} \int_0^L dy [g(y)]^2$. $Q_u=\frac{2}{3}$ and $Q_d=-\frac{1}{3}$ are the up- and down-type quark electric charges in units of the positron charge $e$. The quantities $I_0^{(\prime)}$, $A^{(\prime)}$ and $B^{(\prime)}$ correspond to the loop integrals,  and  they are listed below.
\begin{widetext}
\bea
I_0(t)&=& \frac{i}{(4 \pi)^2}\,\frac{1}{M^2_{KK}}\left(-\frac{1}{t-1}+\frac{\ln (t)}{(t-1)^2} \right)\nn \\
I_0^\prime(t)&=& \frac{i}{(4 \pi)^2}\,\frac{1}{M^4_{KK}}
\left(\frac{1+t}{2t(t-1)^2}-\frac{\ln (t)}{(t-1)^3} \right)\nn  \\
A(t) &=&  B(t)=\frac{i}{(4 \pi)^2}\,\frac{1}{4 M^2_{KK}} \left(\frac{t-3}{(t-1)^2}+\frac{2\ln (t)}{(t-1)^3} \right)\nn \\
A^\prime(t) &=& 2 B^\prime(t)=\frac{i}{(4 \pi)^2}\,\frac{1}{ M^4_{KK}} 
\left(-\frac{t^2-5t-2}{6t(t-1)^3}-\frac{\ln (t)}{(t-1)^4} \right) \label{loopint}  \\
{\bar A}(t) &=&  {\bar B}(t)=\frac{i}{(4 \pi)^2}\,\frac{1}{4 M^2_{KK}} \left(-\frac{3t-1}{(t-1)^2}+\frac{2t^2\ln (t)}{(t-1)^3} \right)\nn \\
{\bar A}^\prime(t) &=& {\bar B}^\prime(t) = \frac{i}{(4 \pi)^2}\,\frac{1}{4 M^4_{KK}} \left(\frac{5t+1}{(t-1)^3}-\frac{2t(2+t)\ln (t)}{(t-1)^4} \right) \,\,\, , \nn 
\eea
with $t=m_F^2/M_{KK}^2$. 
\end{widetext}
%

%\clearpage

\end{document}